\PassOptionsToPackage{unicode}{hyperref}
\PassOptionsToPackage{hyphens}{url}
\PassOptionsToPackage{dvipsnames,svgnames,x11names}{xcolor}
\documentclass[
  12pt]{article}

\usepackage{amsmath,amssymb}
\usepackage{iftex}
\ifPDFTeX
  \usepackage[T1]{fontenc}
  \usepackage[utf8]{inputenc}
  \usepackage{textcomp} 
\else 
  \usepackage{unicode-math}
  \defaultfontfeatures{Scale=MatchLowercase}
  \defaultfontfeatures[\rmfamily]{Ligatures=TeX,Scale=1}
\fi
\usepackage{lmodern}
\ifPDFTeX\else  
\fi
\IfFileExists{upquote.sty}{\usepackage{upquote}}{}
\IfFileExists{microtype.sty}{
  \usepackage[]{microtype}
  \UseMicrotypeSet[protrusion]{basicmath} 
}{}
\makeatletter
\@ifundefined{KOMAClassName}{
  \IfFileExists{parskip.sty}{%
    \usepackage{parskip}
  }{
    \setlength{\parindent}{0pt}
    \setlength{\parskip}{6pt plus 2pt minus 1pt}}
}{
  \KOMAoptions{parskip=half}}
\makeatother
\usepackage{xcolor}
\makeatletter
\ifx\paragraph\undefined\else
  \let\oldparagraph\paragraph
  \renewcommand{\paragraph}{
    \@ifstar
      \xxxParagraphStar
      \xxxParagraphNoStar
  }
  \newcommand{\xxxParagraphStar}[1]{\oldparagraph*{#1}\mbox{}}
  \newcommand{\xxxParagraphNoStar}[1]{\oldparagraph{#1}\mbox{}}
\fi
\ifx\subparagraph\undefined\else
  \let\oldsubparagraph\subparagraph
  \renewcommand{\subparagraph}{
    \@ifstar
      \xxxSubParagraphStar
      \xxxSubParagraphNoStar
  }
  \newcommand{\xxxSubParagraphStar}[1]{\oldsubparagraph*{#1}\mbox{}}
  \newcommand{\xxxSubParagraphNoStar}[1]{\oldsubparagraph{#1}\mbox{}}
\fi
\makeatother

\usepackage{longtable,booktabs,array}
\usepackage{calc} 
\usepackage{etoolbox}
\makeatletter
\patchcmd\longtable{\par}{\if@noskipsec\mbox{}\fi\par}{}{}
\makeatother
\IfFileExists{footnotehyper.sty}{\usepackage{footnotehyper}}{\usepackage{footnote}}
\makesavenoteenv{longtable}
\usepackage{graphicx}
\makeatletter
\def\maxwidth{\ifdim\Gin@nat@width>\linewidth\linewidth\else\Gin@nat@width\fi}
\def\maxheight{\ifdim\Gin@nat@height>\textheight\textheight\else\Gin@nat@height\fi}
\makeatother
\setkeys{Gin}{width=\maxwidth,height=\maxheight,keepaspectratio}
\makeatletter
\def\fps@figure{htbp}
\makeatother

\makeatletter
\@ifpackageloaded{caption}{}{\usepackage{caption}}
\AtBeginDocument{%
\ifdefined\contentsname
  \renewcommand*\contentsname{Table of contents}
\else
  \newcommand\contentsname{Table of contents}
\fi
\ifdefined\listfigurename
  \renewcommand*\listfigurename{List of Figures}
\else
  \newcommand\listfigurename{List of Figures}
\fi
\ifdefined\listtablename
  \renewcommand*\listtablename{List of Tables}
\else
  \newcommand\listtablename{List of Tables}
\fi
\ifdefined\figurename
  \renewcommand*\figurename{Figure}
\else
  \newcommand\figurename{Figure}
\fi
\ifdefined\tablename
  \renewcommand*\tablename{Table}
\else
  \newcommand\tablename{Table}
\fi
}
\@ifpackageloaded{float}{}{\usepackage{float}}
\floatstyle{ruled}
\@ifundefined{c@chapter}{\newfloat{codelisting}{h}{lop}}{\newfloat{codelisting}{h}{lop}[chapter]}
\floatname{codelisting}{Listing}

\makeatother
\makeatletter
\@ifpackageloaded{caption}{}{\usepackage{caption}}
\@ifpackageloaded{subcaption}{}{\usepackage{subcaption}}
\makeatother

\ifLuaTeX
  \usepackage{selnolig}  
\fi
\usepackage[]{natbib}
\usepackage{bookmark}

\IfFileExists{xurl.sty}{\usepackage{xurl}}{} 
\hypersetup{
  pdftitle={Title},
  pdfauthor={Author 1; Author 2},
  pdfkeywords={3 to 6 keywords, that do not appear in the title},
  colorlinks=true,
  linkcolor={blue},
  filecolor={Maroon},
  citecolor={Blue},
  urlcolor={Blue},
  pdfcreator={LaTeX via pandoc}}

\newtheorem{theorem}{Theorem}
\newtheorem{lemma}{Lemma}

\newcommand{\anon}{1}

\usepackage{newunicodechar}
\newunicodechar{∗}{\textasteriskcentered}

\begin{document}

\def\spacingset#1{\renewcommand{\baselinestretch}%
{#1}\small\normalsize} \spacingset{1}


\if1\anon
{
  \title{\bf Finite Population Inference for Factorial Designs and Panel Experiments with Imperfect Compliance}
  \author{Pedro Picchetti\thanks{\textbf{Pedro Picchetti:} Instituto de Economía, PUC Chile. E-mail: pedro.picchetti@uc.cl\\ 
I thank Cristine Pinto, Jonathan Roth, Peter Hull, Sergio Firpo, Luis Alvarez, Sukjin Han, Toru Kitagawa, Vitor Possebom, as well as participants in the Bristol Econometric Study Group 2024, the New York Camp Econometrics XVIII, the Brown Econometrics Seminar and the Insper Students Seminar. I gratefully acknowledges the financial support from FAPESP 2021/13708-8 and FAPESP 2022/13229-5. All errors are my own.\\}\hspace{.2cm}\\
    Institute of Economics, PUC-Chile}
  \maketitle
} \fi

\if0\anon
{
  \bigskip
  \bigskip
  \bigskip
  \begin{center}
    {\LARGE\bf Title}
\end{center}
  \medskip
} \fi

\bigskip
\begin{abstract}
This paper develops a finite population framework for analyzing causal effects in settings with imperfect compliance where multiple treatments affect the outcome of interest. Two prominent examples are factorial designs and panel experiments with imperfect compliance. I define finite population causal effects that capture the relative effectiveness of alternative treatment sequences. I provide nonparametric estimators for a rich class of factorial and dynamic causal effects and derive their finite population distributions as the sample size increases. Monte Carlo simulations illustrate the desirable properties of the estimators. Finally, I use the estimator for causal effects in factorial designs to revisit a famous voter mobilization experiment that analyzes the effects of voting encouragement through phone calls on turnout.
\end{abstract}

\noindent%
\vfill

\newpage
\spacingset{1.8} 

\section{Introduction}

In a seminal paper, \cite{imbensangrist} showed that the instrumental variables (IV) estimand in cross-sectional settings with a binary treatment and a binary instrument can be interpreted as the Local Average Treatment Effect (LATE), defined as the average treatment effect for the subpopulation that has its treatment status shifted by an excluded instrument, the so-called compliers.

However, several applications of IV methods deviate from the canonical setting. For instance, it is common to see researchers using IV methods in settings which the outcome depends on multiple treatments. Prominent examples are factorial designs \citep{Gerber_Green_2000,nick} and panel experiments \citep{stango} with imperfect compliance. In such settings, it is well known that standard IV methods can lead to misleading conclusions regarding causal effects.

In this paper, I develop a novel nonparametric identification approach for causal effects in factorial designs and panel experiments. Using standard instrumental variables assumptions along with a exclusion restriction for treatments, I show that causal effects associated to sequences of treatments (a vector of factors of interest, or the path of treatments taken through time), can be identified by exploiting variations in the sequence of assignments associated to such treatments.

The intuition behind this result is simple: under the treatment exclusion restriction, if we are interested in the effect of a single treatment, we identify its causal effects for compliers of its assignment by exploiting variations in that assignment. If we are interested in the causal effects of a sequence of two treatments, we identify causal effects by exploiting variations in the four possible assignments in a difference-in-differences format. In the case of a sequence of three treatments causal effects are identified using triple-differences, and so on.

My approach takes a purely design-based perspective on uncertainty, which allows for potential outcomes models to remain unspecified. I propose nonparametric estimators for causal effects which are consistent over the randomization distribution and derive their finite population asymptotic distributions. I use the results to propose valid inference procedures, which are potentially conservative in the presence of heterogeneous causal effects.

I conduct Monte Carlo simulation studies to analyze the properties of these estimators. The results show that the estimators exhibit desirable performance in terms of bias and confidence interval coverage, both in the factorial design and the panel experiment settings. Finally, I revisit the work of \cite{nick} and apply the estimators in a real life famous factorial experiment from the political science literature, which studies the effects of voting encouragements through phone calls in turnout.

This paper contributes to several strands of the causal inference literature. First, to the literature on factorial designs with imperfect compliance \citep{chengsmall,Blackwell2017,schochet,Blackwell2023}. More specifically, I rely on the same assumptions as \cite{Blackwell2017} and \cite{Blackwell2023}, but define a new, rich class of causal effects and provide a new nonparametric estimator.

Second, to the literature on dynamic causal effects in the tradition of \cite{ROBINS19861393,ROBINS1987139S} and \cite{murphy}, where potential outcomes in a given period depend on the path of treatments taken until that period. I focus on the case of imperfect compliance and the identification of lag-$p$ dynamic causal effects, defined by \cite{bojinovshephard}. I extend the results of \cite{bojinov} to the case of panel experiments with imperfect compliance. I motivate my approach for the identification of dynamic causal effects by showing that standard IV estimands in general do not have a straightforward causal interpretation in the presence of dynamic causal effects even in the case where the exclusion restriction for treatments hold. See Appendix C for a causal decomposition of the Wald estimand in a setting with two time periods.

Finally, the paper relates to the literature on design-based inference in IV settings (\cite{Imbens_Rubin_2015}; \cite{kang}; \cite{jonashesh}, \cite{peterkyrill}). While these papers focus solely on estimators for IV settings with a single treatment and a single instrument, I focus on settings with multiple treatments and instruments. While Blackwell 2023 also provides finite population inference procedures, this is the first paper that studies design-based inference in panel experiments with imperfect compliance, to the best of my knowledge.

The paper is organized as follows. In Section 2 I introduce the framework for factorial designs and provide the identification results, estimators and its asymptotic properties for this setting. In section 3 I proceed similarly, but focus on the case of panel experiments. In Section 4 I conduct Monte Carlo simulations to analyze the properties of the estimators introduced previously, and in Section 5 I apply the techniques developed in Section 2 to the voter mobilization empirical application. Section 6 concludes. The supplemental appendix contains the proofs of the theorems presented in this paper and the auxiliary lemmas.

\section{Factorial Designs}

\subsection{Framework and Target Parameters}

Consider a setting in which $N$ units are part of an experiment with $K$ binary factors with levels $\left\{0,1\right\}$, where 0 denote the untreated level and 1 denote the treated level. Let $Z_{i,k}$ be an indicator for units assigned to the treated level for factor $k$ and $D_{i,k}$ denote an indicator for units that uptake treatment in factor $k$. Index sets are compactly written as $[N]:=\left\{1,...,N\right\}$ and $[K]:=\left\{1,...,K\right\}$.

Imperfect compliance arises from the fact that not all units assigned to treatment in factor $k$ actually take treatment, and not all units assigned to control remain untreated, that is, for all $k\in[K]$, $D_{i,k}\neq Z_{i,k}$ for some $i\in[N]$.

For an individual $i$, I denote the vectors of treatment and assignments up to factor $k$ respectively as $D_{i,1:k}$ and $Z_{i,1:k}$. Vectors of treatments and assignments that do not contain elements for factors $k-p$ to $k$ are defined as $D_{i,-(k-p:k)}$ and $Z_{i,-(k-p:k)}$. For a given factor $k$, I denote the cross-section of treatments and assignments respectively as $D_{1:N,k}$ and $Z_{1:N,k}$. Collections over individuals and factors are denoted by $D_{1:N,1:K}$ and $Z_{1:N,1:K}$.

Without further restrictions, potential outcomes of unit $i$ are a function of the full sequence of treatments and assignments, $Y_{i}(d_{1:N,1:K},z_{1:N,1:K})$, and potential treatments are a function of the full sequence of assignments, $D_{i,k}(z_{1:N,1:K})$. The next assumptions are invoked for identification.

\textbf{Assumption 1 (No-Spillovers Across Units):} For all $i\in[N]$ and $k\in[K]$,

\begin{align*}
    &Y_{i}=\sum_{(d_{1:K},z_{1:K})\in\left\{ 0,1\right\}^{K\times K}}\mathbf{1}\left\{ D_{i,1:K}=d_{1:K},Z_{i,1:K}=z_{1:K}\right\}Y_{i}(d_{1:K},z_{1:K}),\\&D_{i,k}=\sum_{(z_{1:K})\in\left\{ 0,1\right\}^{K}}\mathbf{1}\left\{ Z_{i,1:K}=z_{1:K}\right\}D_{i,k}(z_{1:K})
\end{align*}

Assumption 1 imposes that the potential outcome of unit $i$ depends only on the treatments and assignments of unit $i$, ruling out the possibility of spillover of both treatments and assignments across units. Assumption 1 is usually referred to as the the Stable Unit Treatment Value Assumption \citep{rubin80}.

One of the fundamental assumptions of IV settings is the exclusion restriction. Its version for factorial designs is stated below, alongside an additional exclusion restriction for the first stage:

\textbf{Assumption 2 (Exclusion Restrictions):} For all $i\in[N]$ and $k\in[K]$, (i) $Y_{i}(d_{1:K},z_{1:K})=Y_{i}(d_{1:K})$ and (ii) $D_{i,k}(z_{1:K})=D_{i,k}(z_{k})$.

The first part of Assumption 2 is the standard exclusion restriction for IV
settings. It states that the sequence of assignments does not affect potential outcomes directly. Assignments only affect potential outcomes to the extent that they affect treatment choices.

The second part of Assumption 2 restricts how assignments affect treatment choices. states that treatment uptake on
factor $k$ only depends on the treatment assignment for factor $k$, not other factors, and is widely invoked in factorial design settings with imperfect compliance \citep{Blackwell2017,Blackwell2023}.

The fundamental behavioral assumption in IV settings is the monotonicity assumption,
which is provided below.

\textbf{Assumption 3 (Monotonicity):} For all $i\in[N]$, $k\in[K]$, $D_{i,k}(1)\geq D_{i,k}(0)$.

Monotonicity states that for each factor, units assigned to treatment ($Z_{i,k} = 1$) are at least as likely to take that treatment as units assigned to control ($Z_{i,k} = 0$). Under Assumption 3, units can be divide into three groups at each period of time defined by how units treatment choice for factor $k$ relates to assignment $k$: Always-takers ($AT_{k}$), Never-takers ($NT_{k}$) and Compliers ($C_{k}$). Note that an individual can be part of a group in a given for a given factor but it does not need to be in the same group through the whole sequence of assignments. Assumptions 2 (ii) and 3 combined imply that there are $3^{K}$ compliance types in an experiment with $K$ factors.

Finally, it is assumed that assignments for all factors are completely randomized, in the sense that they are independent from potential outcomes and treatments, and from assignments to other factors. Let $\mathcal{F}_{i}:=\left\{Y_{i}(.), D_{i,1:K}(.)\right\}$. The assumption is formalized below:

\textbf{Assumption 4 (Individualistic Assignments):}For all $i\in[N]$ and $k\in[K]$,

\begin{equation*}
    \mathbb{P}\left ( Z_{i,k}|Z_{-i,k},Z_{i,-k},\mathcal{F}_{1:N} \right )=\mathbb{P}\left ( Z_{i,k} \right )
\end{equation*}

Assumption 4 states that the assignment for factor $k$ for unit $i$ is independent from the assignment of other units, the assignment for other factors and potential quantities. It is readily satisfied in completely randomized experiments (where $\mathbb{P}\left ( Z_{i,k}|Z_{-i,k},Z_{i,-k},\mathcal{F}_{1:N} \right )=c_{k}\in \left(0,1\right)$ for all $i\in[N]$), but it also allows for more complex assignment mechanisms.

In the factorial design setting, causal effects are the differences in potential outcomes for unit $i$ associated to different sequences of treatments, which are defined as $\tau_{i}(d_{1:K},\tilde{d}_{1:K})=Y_{i}(d_{1:K})-Y_{i}(\tilde{d}_{1:K})$.

The number of potential outcomes grows exponentially with the number of factors. For the sake of tractability, I focus on the $k-p$ to $k$ joint causal effect:

\begin{equation*}
    \tau_{i,k-p:k}(\textbf{d},\tilde{\textbf{d}})
:=Y_{i}(d^{obs}_{i,-(k-p:k)},\textbf{d})-Y_{i}(d^{obs}_{i,-(k-p:k)},\tilde{\textbf{d}})
\end{equation*}

\noindent which can be interpreted as the causal effect of taking a sequence of treatments from factor $k-p$ to $k$ versus an alternative sequence, keeping the sequence of all other factors fixed.

In settings with imperfect compliance, the average causal effects are only identifiable for the group of compliers. I define a target parameters for the group of compliers of the sequence. The local $k-p$ to $k$ causal effect

\begin{equation*}
    \overline{\tau}_{C_{k-p:k}}(\textbf{d},\tilde{\textbf{d}})=\frac{1}{\left | C_{k-p:k} \right |}\sum_{i\in C_{k-p:k}}\tau_{i,k-p:k}(\textbf{d},\tilde{\textbf{d}})
\end{equation*}

\noindent which is the $k-p$ to $k$ causal effect for individuals that are compliers for all treatments considered within the sequence of interest.

\subsection{Identification}

I show that joint causal effects can be identified after potential outcomes associated to different sequences of treatment are identified separately. Define the local $k-p$ to $k$ response function as 

\begin{equation*}
    m_{k-p:k}(\textbf{d})=\frac{1}{\left | C_{k-p:k}\right |}\sum_{i\in C_{k-p:k}}Y_{i}(d^{obs}_{i,-(k-p:k)},\textbf{d)}
\end{equation*}

\noindent for any $\textbf{d}\in\left\{0,1\right\}^{p+1}$. I show that local response functions can be identified for all combinations of assignments. The local causal effect is then generated by taking the difference between two local response functions: $\tau_{C_{k-p:k}}(\textbf{d},\tilde{\textbf{d}})=m_{k-p:k}(\textbf{d})-m_{k-p:k}(\tilde{\textbf{d}})$.

Before introducing the estimand, I defined the \textit{adapted propensity scores} from factors $k-p$ to $k$ as $\pi_{i,k-p:k}(\textbf{z})=\mathbb{P}\left(Z_{i,k-p:k}=\textbf{z}\right)$, which can be computed along the observed assignment sequences. The fundamental object used for identification is the Horvitz-Thompson estimand. For a generic random variable $R_{i,t}$, the Horvitz-Thompson associated to the assignment sequence $\textbf{z}\in\left\{0,1\right\}^{p+1}$ is

\begin{equation*}
    \mathbb{E}\left [ \frac{1}{N}\sum_{i=1}^{N}\frac{R_{i}\mathbf{1}\left\{ Z_{i,k-p:k}=\textbf{z}\right\}}{\pi_{i,k-p:k}(\textbf{z})} |\mathcal{F}_{1:N}\right ]
\end{equation*}

The Horvitz-Thompson-type quantities are combined in what I call a multiple-differences format. Below, I define an expression for the general multiple difference in means across assignment paths, which I will refer to hereafter as the $p+1$ Horvitz-Thompson multiple-difference ($\Delta^{p+1}_{HT}$).

\textbf{Definition ($\Delta^{p+1}_{HT}$):} Define the $p+1$-th difference of Horvitz-Thompson estimands across assignment sequences from period $k-p$ to period $k$ of a random variable $R$ as is such that for $p\geq 0$,

\begin{align*}
    &\Delta^{p+1}_{HT}\left ( \mathbb{E}\left [ \frac{1}{N}\sum_{i=1}^{N}\frac{R_{i}\mathbf{1}\left\{Z_{i,k-p:k}=\textbf{z}\right\}}{\pi_{i,k-p:k}(\textbf{z})}|\mathcal{F}_{1:N,-(k-p:k)} \right ] \right )=\\&\Delta^{p}_{HT}\left ( \mathbb{E}\left [ \frac{1}{N}\sum_{i=1}^{N}\frac{R_{i}\mathbf{1}\left\{Z_{i,k-p:k}=1,\textbf{z}\right\}}{\pi_{i,k-p:k}(1,\textbf{z})}|\mathcal{F}_{1:N,-(k-p:)} \right ] \right )\\&-\Delta^{p}_{HT}\left ( \mathbb{E}\left [ \frac{1}{N}\sum_{i=1}^{N}\frac{R_{i}\mathbf{1}\left\{Z_{i,k-p:k}=0,\textbf{z}\right\}}{\pi_{i,k-p:k}(0,\textbf{z})}|\mathcal{F}_{1:N,-(k-p:k)} \right ] \right )
\end{align*}

When $p=0$, the operator is simply a difference in means:

\begin{align*}
    &\Delta^{0}_{HT}\left ( \mathbb{E}\left [ \frac{1}{N}\sum_{i=1}^{N}\frac{R_{i}\mathbf{1}\left\{Z_{i,k}=\textbf{z} \right\}}{\pi_{i,k}(\textbf{z})}|\mathcal{F}_{1:N,-k} \right ] \right )\\&=\mathbb{E}\left [ \frac{1}{N}\sum_{i=1}^{N}\frac{R_{i}\mathbf{1}\left\{Z_{i,k}=1 \right\}}{\pi_{i,k}(1)}|\mathcal{F}_{1:N,-k} \right ]-\mathbb{E}\left [ \frac{1}{N}\sum_{i=1}^{N}\frac{R_{i}\mathbf{1}\left\{Z_{i,k}=0 \right\}}{\pi_{i,k}(0)}|\mathcal{F}_{1:N,-k} \right ]
\end{align*}

When $p=1$, the multiple differences operator takes the form of difference-in-differences:

{\footnotesize \begin{align*}
    &\Delta^{1}_{HT}\left ( \mathbb{E}\left [ \frac{1}{N}\sum_{i=1}^{N}\frac{R_{i}\mathbf{1}\left\{Z_{i,k-1:k}=\textbf{z} \right\}}{\pi_{i,k-1:k}(\textbf{z})}|\mathcal{F}_{1:N,-(k-1:k)} \right ] \right )\\&=\Delta^{0}_{HT}\left ( \mathbb{E}\left [ \frac{1}{N}\sum_{i=1}^{N}\frac{R_{i}\mathbf{1}\left\{Z_{i,k-1:k}=(1,\textbf{z}) \right\}}{\pi_{i,k-1:k}(1,\textbf{z})}|\mathcal{F}_{1:N,-(k-1:k)} \right ] \right )\\&-\Delta^{0}_{HT}\left ( \mathbb{E}\left [ \frac{1}{N}\sum_{i=1}^{N}\frac{R_{i}\mathbf{1}\left\{Z_{i,k-1:k}=(0,\textbf{z}) \right\}}{\pi_{i,k-1:k}(0,\textbf{z})}|\mathcal{F}_{1:N,-(k-1:k)} \right ] \right )\\&=\mathbb{E}\left [ \frac{1}{N}\sum_{i=1}^{N}\frac{R_{i}\mathbf{1}\left\{Z_{i,k-1:k}=(1,1) \right\}}{\pi_{i,k-1:k}(1,1)}|\mathcal{F}_{1:N,-(k-1:k)} \right ]-\mathbb{E}\left [ \frac{1}{N}\sum_{i=1}^{N}\frac{R_{i}\mathbf{1}\left\{Z_{i,k-1:k}=(1,0) \right\}}{\pi_{i,k-1:k}(1,0)}|\mathcal{F}_{1:N,-(k-1:k)} \right ]\\&-\left(\mathbb{E}\left [ \frac{1}{N}\sum_{i=1}^{N}\frac{R_{i}\mathbf{1}\left\{Z_{i,k-1:k}=(0,1) \right\}}{\pi_{i,k-1:k}(0,1)}|\mathcal{F}_{1:N,-(k-1)} \right ]-\mathbb{E}\left [ \frac{1}{N}\sum_{i=1}^{N}\frac{R_{i}\mathbf{1}\left\{Z_{i,k-1:k}=(0,0) \right\}}{\pi_{i,k-1:k}(0,0)}|\mathcal{F}_{1:N,-(k-1:k)} \right ]\right)
\end{align*}}

When $p=2$ the operator takes the form of a triple-difference, and so on. Thus, the $p+1$ difference can be used to exploit all possible variations in assignment paths from $k$ to $k-p$, while keeping the rest of the assignments fixed.

In order to use the Horvitz-Thompson multiple-difference estimands, one must assume that there is common support for the adapted propensity score for all possible sequences of factors.

\textbf{Assumption 5:} There exists $C^{L}<C^{U}\in(0,1)$ such that or all $i\in[N]$, $k\in[K]$,  $C^{L}<\pi_{i,k-p:k}(\textbf{z})<C^{U}$ for all $\textbf{z}\in\left \{ 0,1 \right \}^{p+1}$.

The theorem below shows that causal effects associated to a single factor can be identified using a Wald-type estimand and that local response functions associated to a treatment path from $k-p$ to $k$ can be identified exploiting variations in the path of assignments from $k-p$ to $k$ in the multiple-differences format.

\begin{theorem}
    Under Assumptions 1-5,

    \begin{equation*}
        \frac{\mathbb{E}\left [\frac{1}{N} \sum_{i=1}^{N}\frac{Z_{i,k}Y_{i}}{\pi_{i,k}(1)}-\frac{1}{N}\sum_{i=1}^{N}\frac{(1-Z_{i,k})Y_{i}}{\pi_{i,k}(0)}|\mathcal{F}_{1:N,-k} \right ]}{\mathbb{E}\left [\frac{1}{N} \sum_{i=1}^{N}\frac{Z_{i,k}D_{i,k}}{\pi_{i,k}(1)}-\frac{1}{N}\sum_{i=1}^{N}\frac{(1-Z_{i,k})D_{i,k}}{\pi_{i,k}(0)}|\mathcal{F}_{1:N,-k} \right ]}=\tau_{C_{k}}(1,0)
    \end{equation*}

    and

    \begin{equation*}
        \frac{\Delta^{p+1}_{HT}\left ( \mathbb{E}\left [ \frac{1}{N}\sum_{i=1}^{N}\frac{Y_{i}\mathbf{1}\left\{D_{i,k-p:k}=\textbf{d} \right\}\mathbf{1}\left\{Z_{i,k-p:k}=\textbf{z} \right\}}{\pi_{i,k-p:k}(\textbf{z})}|\mathcal{F}_{1:N,-(k-p:k)} \right ] \right )}{\Delta^{p+1}_{HT}\left ( \mathbb{E}\left [ \frac{1}{N}\sum_{i=1}^{N}\frac{\mathbf{1}\left\{D_{i,k-p:k}=\textbf{d} \right\}\mathbf{1}\left\{Z_{i,k-p:k}=\textbf{z} \right\}}{\pi_{i,k-p:k}(\textbf{z})}|\mathcal{F}_{1:N,-(k-p:k)} \right ] \right )}=m_{k-p:k}(\textbf{d})
    \end{equation*}
\noindent for any $\textbf{d}\in\left\{0,1\right\}^{p+1}$.
\end{theorem}

Theorem 1 shows that the causal effect of factor $k$ is identified by a standard two-stage Horvitz-Thompson estimand. In order to identify causal effects associated to multiple factors, local response functions are identified separately by exploiting variation in the sequence of assignments that affect the factors of interest in the multiple-differences format.

\subsection{Estimation and Inference}

In this section, I study the asymptotic properties of the estimator corresponding to the estimands introduced in Theorem 1.

I first derive the asymptotic properties of the factor $k$ causal effect estimator over the randomization distribution and then proceed with the asymptotic properties of the estimator for general $k-p$ to $k$ causal effects.

When it comes to the case where we are interested in the causal effect of a single treatment, potential outcomes do not need to be identified separately. I propose a simple two-stage Horvitz-Thompson (HT) estimator, which is built using the adapted propensity score.

The nonparametric estimator for $\tau_{i,k}(1,0)$ is

\begin{equation*}
    \widehat{\tau}_{i,k}(1,0)=\frac{\widehat{\tau}_{i,k}^{RF}(1,0)}{\widehat{\tau}_{i,k}^{FS}(1,0)}
\end{equation*}

\noindent where

\vspace{-10mm}

 \begin{align*}
     &\widehat{\tau}_{i,k}^{RF}(1,0)=\frac{Z_{i,k}Y_{i}}{\pi_{i,k}(1)}-\frac{(1-Z_{i,k})Y_{i}}{\pi_{i,k}(0)},\\&\widehat{\tau}_{i,k}^{FS}(1,0)=\frac{Z_{i,k}D_{i,k}}{\pi_{i,k}(1)}-\frac{(1-Z_{i,k})D_{i,k}}{\pi_{i,k}(0)}
 \end{align*}

The local factor $k$ causal effect is then estimated by plugging in $\widehat{\tau}_{i,k}(1,0)$:

\begin{equation*}
    \widehat{\overline{\tau}}_{C_{k}}(1,0)=\frac{\frac{1}{N}\sum_{i=1}^{N}\widehat{\tau}_{i,k}^{RF}(1,0)}{\frac{1}{N}\sum_{i=1}^{N}\widehat{\tau}_{i,k}^{FS}(1,0)}
\end{equation*}

Theorem 2 shows that the estimator is consistent over the randomization distribution, and asymptotically normal as the population size grows larger.

\begin{theorem}
    Suppose that potential outcomes are bounded and Assumptions 1-5 hold. Then, 

\begin{equation*}
    \frac{\sqrt{N}\left \{ \widehat{\overline{\tau}}_{C_{k}}(1,0)-\overline{\tau}_{C_{k}}(1,0) \right \}}{\sigma_{k}(1,0)}\overset{d}{\rightarrow} \mathcal{N}(0,1),\ as\ N\rightarrow\infty
\end{equation*}

\noindent where $\sigma_{k}(1,0)$ is defined in the Appendix.
\end{theorem}

The variances of $\widehat{\overline{\tau}}_{C_{k}}(1,0)$ is the appropriate averages of the variance of $\widehat{\overline{\tau}}_{i,k}(1,0)$, which is generally not estimable as it depends on individual potential outcomes and potential treatments under both treatment and counterfactual. However, Lemma 1 in Appendix B shows that the variance of the reduced form and the first stage are bounded from above by a term that is estimable.

For hypothesis testing, I propose the estimation of a conservative \cite{bloom} confidence interval, built with estimates of the upper bound of the variance of the reduced form, and the square of the estimate of the first stage.

The upper bound for the variance of the estimator of the reduced form $\widehat{\tau}_{i,k}^{RF}(1,0)$ is

\begin{equation*}
    \left ( \gamma_{i,k}^{RF}(1,0) \right )^{2}=\frac{Y_{i}(d^{obs}_{i,-k},D_{i,k}(1))^{2}}{\pi_{i,k}(1)}+\frac{Y_{i}(d^{obs}_{i,-k},D_{i,k}(0))^{2}}{\pi_{i,k}(0)}
\end{equation*}

and it can be consistently estimated by $\left ( \widehat{\gamma}_{i,k}^{RF}(1,0) \right )^{2}=\frac{Y_{i}^{2}\left\{ Z_{i,k}+(1-Z_{ik})\right\}}{\pi_{i,k}(Z_{i,k})^{2}}$ and plugged-in for the estimate of the confidence interval. The resulting $1-\alpha$ confidence intervals for the local factor $k$ causal effect is

\begin{equation*}
    \widehat{\overline{\tau}}_{C_{k}}(1,0)\pm z_{1-\alpha/2}\sqrt{\left \{ \frac{\frac{1}{N^{2}}\sum_{i=1}^{N}\left ( \gamma_{i,k}^{RF}(1,0) \right )^{2}}{\left ( \frac{1}{N}\sum_{i=1}^{N}\widehat{\tau}_{i,k}^{FS}(1,0) \right )^{2}} \right \}}
\end{equation*}

\cite{bloom} intervals exhibit good performance in terms of coverage rates for compliance rates greater than 10\%. See \cite{kang} for a thorough discussion about inference using IV in cross-sectional settings with a single binary factor.

For the general $k-p$ to $k$ local causal effect, I propose the separate estimation of local response functions through a multiple-differences Horvitz-Thompson type of estimator. That is, $\widehat{\tau}_{i,k-p:k}(\textbf{d},\tilde{\textbf{d}})=\widehat{m}_{i,k-p:k}(\textbf{d})-\widehat{m}_{i,k-p:k}(\tilde{\textbf{d}})$, where $\widehat{m}_{i,k-p:k}(\textbf{d})=\frac{\widehat{m}_{i,k-p:k}^{RF}(\textbf{d})}{\widehat{m}_{i,k-p:k}^{FS}(\textbf{d})}$ with

 \begin{align*}
     & \widehat{m}_{i,k-p:k}^{RF}(\textbf{d})=\Delta^{p+1}\left ( \frac{\mathbf{1}\left \{ Z_{i,k-p:k}=\textbf{z} \right \}\mathbf{1}\left \{ D_{i,k-p:k}=\textbf{d} \right \}Y_{i}}{\pi_{i,k-p:k}(\textbf{z})} \right ),\\& \widehat{m}_{i,k-p:k}^{FS}(\textbf{d})=\Delta^{p+1}\left ( \frac{\mathbf{1}\left \{ Z_{i,k-p:k}=\textbf{z} \right \}\mathbf{1}\left \{ D_{i,k-p:k}=\textbf{d} \right \}}{\pi_{i,k-p:k}(\textbf{z})} \right )
 \end{align*}

Plugging the unit $i$ estimates for the $k-p$ to $k$ local response function leads to the estimate of the average local response function, which is 

\begin{equation*}
    \widehat{m}_{k-p:k}(\textbf{d})=\frac{\frac{1}{N}\sum_{i=1}^{N}\widehat{m}_{i,k-p:k}^{RF}(\textbf{d})}{\frac{1}{N}\sum_{i=1}^{N}\widehat{m}_{i,k-p:k}^{FS}(\textbf{d})}
\end{equation*}

Estimates for local $k-p$ to $k$ causal effects are generated by the difference of the estimates for the response functions associated to alternative treatment paths. Theorem 3 shows that the estimators are consistent over the randomization distribution, and asymptotically normal as the population size grows larger.

\begin{theorem}
    Suppose that potential outcomes are bounded and Assumptions 1-5 hold. Then, 

\begin{equation*}
    \frac{\sqrt{N}\left \{ \widehat{\overline{\tau}}_{C_{k-p:k}}(\textbf{d},\tilde{\textbf{d}})-\overline{\tau}_{C_{k-p:k}}(\textbf{d},\tilde{\textbf{d}}) \right \}}{\sigma_{k-p:k}(\textbf{d},\tilde{\textbf{d}})}\overset{d}{\rightarrow} \mathcal{N}(0,1),\ as\ N\rightarrow\infty
\end{equation*}

where $\sigma_{k-p:k}(\textbf{d},\tilde{\textbf{d}})$ is defined in the Appendix.
\end{theorem}

For hypothesis testing, conservative \cite{bloom} confidence intervals are constructed using the upper bound for the variance of the reduced forms for the response functions and estimates of the first stage for the response functions.

The upper bound for the variance of the estimator of the reduced form is

\vspace{-10mm}

\begin{align*}
   & \left ( \gamma_{i,k-p:k}^{RF}(\textbf{d}) \right )^{2}+\left ( \gamma_{i,k-p:k}^{RF}(\tilde{\textbf{d}}) \right )^{2}\\&=\sum_{\textbf{z}\in\left \{ 0,1 \right \}^{p+1}}\frac{\left ( Y_{i}(d^{obs}_{i,-k:p},\textbf{d})\mathbf{1}\left \{ D_{i,k-p:k}(\textbf{z})=\textbf{d} \right \} \right )^{2}}{\pi_{i,k-p:k}(\textbf{z})}+\sum_{\textbf{z}\in\left \{ 0,1 \right \}^{p+1}}\frac{\left ( Y_{i}(d^{obs}_{i,-k:p},\tilde{\textbf{d}})\mathbf{1}\left \{ D_{i,k-p:k}(\textbf{z})=\tilde{\textbf{d}} \right \} \right )^{2}}{\pi_{i,k-p:k}(\textbf{z})}
\end{align*}

and it can be consistently estimated and subsequently plugged-in to construct the confidence interval.

In Section 4.1, I conduct Monte Carlo simulations to study the asymptotic properties of these estimators.

\section{Panel Experiments}

\subsection{Framework and Assumptions}

Consider a balanced panel in which $N$ units are observed over $T$ periods of time. For each unit $i\in[N]$ and time period $t\in[T]$, we observe a binary instrumental variable $Z_{i,t}\in\left \{ 0,1 \right \}$, a binary treatment status $D_{i,t}\in\left \{ 0,1 \right \}$ and a real-valued scalar outcome $Y_{i,t}$. Let $\mathcal{F}_{1:N,1:t}$ denote the filtration generated by $Z_{1:N,1:t}$ and the panel of potential quantities. Note that since we are taking a design-based approach, conditioning on $Z_{i,1:t}$ is the same as conditioning on $Z_{i,1:t}$ and the potential quantities associated to such assignment path.

Without further restrictions, potential outcomes of unit $i$ in period $t$ are a function of the full panel of treatments and assignments, $Y_{i,t}(d_{1:N,1:T},z_{1:N,1:T})$, and potential treatments are a function of the full panel of assignments, $D_{i,t}(z_{1:N,1:T})$. The next assumptions are invoked for identification.

\textbf{Assumption 6 (No-Spillovers and No-Anticipation):} For all $i\in[N]$, $t\in[N]$, 

\vspace{-10mm}

\begin{align*}
    & Y_{i,t}=\sum_{(d_{1:t},z_{1:t})\in\left \{ 0,1 \right \}^{t\times t}}\mathbf{1}\left \{D_{i,1:t}=d_{1:t},Z_{i,1:t}=z_{1:t}  \right \}Y_{i,t}(d_{1:t},z_{1:t})\\
    &  D_{i,t}=\sum_{z_{1:t}\in\left \{ 0,1 \right \}^{t}}\mathbf{1}\left \{Z_{i,1:t}=z_{1:t}  \right \}D_{i,t}(z_{1:t})
\end{align*}

Assumption 6 imposes that the potential outcome of unit $i$ in period $t$ depends only on the treatment and assignment paths of unit $i$ until period $t$, ruling out the possibility of spillover of both treatments and assignments across units, as well as future treatments affecting past potential outcomes. It also imposes that the potential treatment from unit $i$ in period $t$ depends only on the assignment paths of unit $i$ until period $t$. To put it shortly, Assumption 6 imposes both SUTVA and no-anticipation.

The dynamic version of the exclusion restriction is stated below, alongside an additional exclusion restriction for the first stage.

\textbf{Assumption 7 (Exclusion Restrictions):} For all $i\in[N]$ and $t\in[T]$, (i) $Y_{i,t}(d_{1:t},z_{1:t})=Y_{i,t}(d_{1:t})$ and (ii) $D_{i,t}(z_{1:t})=D_{i,t}(z_{t})$.

The first part of Assumption 7 is the standard exclusion restriction for dynamic IV settings. It states that the path of assignments does not affect potential outcomes directly. Assignments only affect potential outcomes to the extent that they affect treatment choices.

The second part of Assumption 7 adapts the treatment exclusion restriction from factorial designs to the panel experiment setting. It imposes that potential treatments in period $t$ depend only on the instruments in period $t$, formalizing the intuition that assignment is “targeted” towards a single treatment.

\textbf{Assumption 8 (Monotonicity)}: For all $i\in[N]$ and $t\in[T]$, $D_{i,t}(1)\geq D_{i,t}(0)$.

Monotonicity states that at each period units assigned to treatment ($Z_{i,t}=1$) are at least as likely to take treatment as units assigned to control ($Z_{i,t}=0$). Under Assumption 3, units can be divide into three groups at each period of time defined by how units treatment choice in period t relates to treatment assignment in period t: Always-takers ($AT_{t}$), Never-takes ($NT_{t}$) and Compliers ($C_{t}$). Note that an individual can be part of a group in a given period, it does not need to be in the same group through the whole path of assignments.

Assumption 9 is the standard exogeneity assumption for assignments in panel experiments

\textbf{Assumption 9 (Individualistic and sequentially randomized assignment):} For all $i\in[N]$, $t\in[T]$ and $z_{1:N,1:t-1}\in \left\{ 0,1\right\}^{N\times (t-1)}$, 

\begin{equation*}
    \mathbb{P}\left ( Z_{i,t}|Z_{-i,t},\mathcal{F}_{1:N,1:t-1} \right )=\mathbb{P}\left ( Z_{i,t}|Z_{i,1:t-1}=z_{1:t-1},D_{i,1:t-1}(z_{1:t-1}),Y_{i,1:t-1}(D_{i,1:t-1}(z_{1:t-1})) \right )
\end{equation*}

Assumption 9 is the exogeneity assumption from \cite{bojinovshephard} and \cite{bojinov}, and imposes that conditional on its own past assignments, treatments and outcomes, the assignment for unit $i$ at time $t$ is independent of the past assignments and outcomes of all other units as well as all other contemporaneous assignments.

Define the \textit{adapted propensity score} for the panel experiment setting as

\begin{equation*}
    \pi_{i,t-p:t}(\textbf{z})=\mathbb{P}\left ( Z_{i,t-p:t}=\textbf{z}|Z_{i,1:t-p-1},D_{i,1:t}(Z_{i,1:t-p-1},\textbf{z}),Y_{i,t}(D_{i,1:t}(Z_{i,1:t-p-1},\textbf{z})) \right )
\end{equation*}

The common support assumption under which it is a valid building block for the estimands is stated below:

\textbf{Assumption 10 (Common Support):} There exists $C^{L}<C^{U}\in(0,1)$ such that or all $i\in[N]$, $t\in[T]$,  $C^{L}<\pi_{i,t-p:t}(\textbf{z})<C^{U}$ for all $\textbf{z}\in\left \{ 0,1 \right \}^{p+1}$.

Define the dynamic causal effect of a treatment path versus an alternative treatment path in period $t$ as $\tau_{i,t}(d_{i,1:t},\tilde{d}_{i,1:t})=Y_{i,t}(d_{i,1:t})-Y_{i,t}(\tilde{d}_{i,1:t})$.

The number of potential outcomes grows exponentially with the periods of time. For the sake of tractability, it is common to focus on lag-p dynamic causal effect as defined in \cite{bojinov}. For $0\leq p\leq t$, and $\textbf{d},\tilde{\textbf{d}}\in\left \{ 0,1 \right \}^{p+1}$, the lag-p dynamic causal effect is defined as

\vspace{-10mm}

\begin{equation*}
    \tau_{i,t}(\textbf{d},\tilde{\textbf{d}};p)
:=Y_{i,t}(d^{obs}_{i,1:t-p-1},\textbf{d})-Y_{i,t}(d^{obs}_{i,1:t-p-1},\tilde{\textbf{d}})
\end{equation*}

\noindent which can be interpreted as the causal effect of taking a treatment path from period $t-p$ to $p$ versus an alternative path, keeping the path until period $t-p-1$ fixed.

The local time $t$ lag-$p$ dynamic causal effects and the time $t$ local lag-$p$ response functions are defined, respectively, as

\begin{align*}
    &\overline{\tau}_{C_{t-p:t},t}(\textbf{d},\tilde{\textbf{d}};p)=\frac{1}{\left |C_{t-p:t} \right |}\sum_{i\in C_{t-p:t}}\tau_{i,t}(\textbf{d},\tilde{\textbf{d}};p),\\&m_{t-p:t}(\textbf{d})=\frac{1}{\left | C_{t-p:t}\right |}\sum_{i\in C_{t-p:t}}Y_{i,t}(d^{obs}_{i,1:t-p-1},\textbf{d})
\end{align*}

Next, I show that the quantities are also identified using a multiple-differences Wald type of estimand.

\subsection{Identification}

Similar to the case of factorial designs, in panel experiments with imperfect compliance where exclusion restrictions for the treatments hold, potential outcomes associated to sequences of treatments are identified by exploiting variations in the corresponding sequence of assignments using the multiple-differences operator for Horvitz-Thompson quantities. Theorem 4 shows that the local lag-0 dynamic causal effect is identified under a two-stage Horvitz-Thompson estimand, and that local lag-$p$ response functions are identified using the multiple-differences approach.

\begin{theorem}
    Under Assumptions 6-10,

    \begin{equation*}
        \frac{\mathbb{E}\left [\frac{1}{N} \sum_{i=1}^{N}\frac{Z_{i,t}Y_{i,t}}{\pi_{i,t}(1)}-\frac{1}{N}\sum_{i=1}^{N}\frac{(1-Z_{i,t})Y_{i,t}}{\pi_{i,t}(0)}|\mathcal{F}_{1:N,t-1} \right ]}{\mathbb{E}\left [\frac{1}{N} \sum_{i=1}^{N}\frac{Z_{i,t}D_{i,t}}{\pi_{i,t}(1)}-\frac{1}{N}\sum_{i=1}^{N}\frac{(1-Z_{i,t})D_{i,t}}{\pi_{i,t}(0)}|\mathcal{F}_{1:N,t-1} \right ]}=\tau_{C_{t},t}(1,0;0)
    \end{equation*}

    and

    \begin{equation*}
        \frac{\Delta^{p+1}_{HT}\left ( \mathbb{E}\left [ \frac{1}{N}\sum_{i=1}^{N}\frac{Y_{i,t}\mathbf{1}\left\{D_{i,t-p:t}=\textbf{d} \right\}\mathbf{1}\left\{Z_{i,t-p:t}=\textbf{z} \right\}}{\pi_{i,t-p:t}(\textbf{z})}|\mathcal{F}_{1:N,t-p-1} \right ] \right )}{\Delta^{p+1}_{HT}\left ( \mathbb{E}\left [ \frac{1}{N}\sum_{i=1}^{N}\frac{\mathbf{1}\left\{D_{i,t-p:t}=\textbf{d} \right\}\mathbf{1}\left\{Z_{i,t-p:t}=\textbf{z} \right\}}{\pi_{i,t-p:t}(\textbf{z})}|\mathcal{F}_{1:N,t-p-1} \right ] \right )}=m_{t-p:t}(\textbf{d})
    \end{equation*}
\noindent for any $\textbf{d}\in\left\{0,1\right\}^{p+1}$.
\end{theorem}

Theorem 4 shows that the mechanics for identifying dynamic causal effects and local lag-$p$ response functions is the same as the one for identifying causal effects and local response functions in factorial designs. That is because we assume the treatment exclusion restriction holds in both settings. Ultimately, the multiple-difference approach works because compliance types assignment-specific (factor-specific in the factorial design and period-specific in the panel experiment design), and thus shifts in an assignment identifies potential quantities for individuals who comply to that assignment while keeping the response fixed for all other treatments. Therefore, the multiple-differences approach allows for potential quantities for compliers of a sequence of assignments to be sequentially identified by sequentially exploring variations in such sequence of assignments while keeping other assignments fixed.

\subsection{Estimation and Inference}

The estimator procedure for local dynamic causal effects and local dynamic response functions is similar to the one proposed in Section 2.3. The nonparametric estimator for $\tau_{i,t}(1,0;0)$ is

\begin{equation*}
    \widehat{\tau}_{i,t}(1,0;0)=\frac{\widehat{\tau}_{i,t}^{RF}(1,0;0)}{\widehat{\tau}_{i,t}^{FS}(1,0;0)}
\end{equation*}

\noindent where

\vspace{-10mm}

 \begin{align*}
     \widehat{\tau}_{i,t}^{RF}(1,0;0)=\frac{Z_{i,t}Y_{i,t}}{\pi_{i,t}(1)}-\frac{(1-Z_{i,t})Y_{i,t}}{\pi_{i,t}(0)}\\\widehat{\tau}_{i,t}^{FS}(1,0;0)=\frac{Z_{i,t}D_{i,t}}{\pi_{i,t}(1)}-\frac{(1-Z_{i,t})D_{i,t}}{\pi_{i,t}(0)}
 \end{align*}

The time-t lag-0 dynamic causal effect can be estimated by plugging in $\widehat{\tau}_{i,t}(1,0;0)$:

\begin{equation*}
    \widehat{\overline{\tau}}_{C_{t},t}(1,0;0)=\frac{\frac{1}{N}\sum_{i=1}^{N}\widehat{\tau}_{i,t}^{RF}(1,0;0)}{\frac{1}{N}\sum_{i=1}^{N}\widehat{\tau}_{i,t}^{FS}(1,0;0)}
\end{equation*}

Theorem 5 shows that the estimator is consistent over the randomization distribution, and asymptotically normal as the population size grows larger.

\begin{theorem}
     Suppose that potential outcomes are bounded. Under Assumptions 6-10,

     \begin{equation*}
    \frac{\sqrt{N}\left \{ \widehat{\overline{\tau}}_{C_{t},t}(1,0;0)-\overline{\tau}_{C_{t},t}(1,0;0) \right \}}{\sigma_{t}(1,0;0)}\overset{d}{\rightarrow} \mathcal{N}(0,1),\ as\ N\rightarrow\infty
\end{equation*}

\noindent where $\sigma_{t}(1,0;0)$ is defined in Section 5 of Appendix A.

\end{theorem}

The variance of $\widehat{\overline{\tau}}_{C_{t},t}(1,0;0)$ is the appropriate averages of the variance of $\widehat{\overline{\tau}}_{i,t}(1,0;0)$, which are generally not estimable as they depends on individual potential outcomes and potential treatments under both treatment and counterfactual. However, Lemma 4 in Appendix B shows that the variance of the reduced form and the first stage are bounded from above by a term that is estimable.

For the general lag-$p$ dynamic causal effect, I propose the separate estimation of lag-$p$ dynamic response functions through a multiple-differences Horvitz-Thompson type of estimator. That is, $\widehat{\tau}_{i,t}(\textbf{d},\tilde{\textbf{d}};p)=\widehat{m}_{i,t}(\textbf{d})-\widehat{m}_{i,t}(\tilde{\textbf{d}})$, where $\widehat{m}_{i,t}(\textbf{d})=\frac{\widehat{m}_{i,t}^{RF}(\textbf{d})}{\widehat{m}_{i,t}^{FS}(\textbf{d})}$ with

 \begin{align*}
     & \widehat{m}_{i,t}^{RF}(\textbf{d})=\Delta^{p+1}\left ( \frac{\mathbf{1}\left \{ Z_{i,t-p:t}=z_{t-p:t} \right \}\mathbf{1}\left \{ D_{i,t-p:t}=\textbf{d} \right \}Y_{i,t}}{\widehat{\pi}_{i,t-p:t}(z_{t-p:t})} \right )\\& \widehat{m}_{i,t}^{FS}(\textbf{d})=\Delta^{p+1}\left ( \frac{\mathbf{1}\left \{ Z_{i,t-p:t}=z_{t-p:t} \right \}\mathbf{1}\left \{ D_{i,t-p:t}=\textbf{d} \right \}}{\widehat{\pi}_{i,t-p:t}(z_{t-p:t})} \right )
 \end{align*}

Plugging the unit $i$, period $t$ estimates for the lag-$p$ local response function as leads to estimates of the time-$t$ local lag-$p$  response function:

\vspace{-10mm}

\begin{equation*}
    \widehat{m}_{t}(\textbf{d})=\frac{\frac{1}{N}\sum_{i=1}^{N}\widehat{m}_{i,t}^{RF}(\textbf{d})}{\frac{1}{N}\sum_{i=1}^{N}\widehat{m}_{i,t}^{FS}(\textbf{d})}
\end{equation*}

The appropriate lag-p dynamic causal effects are generated by the difference of the estimates for the response functions. Theorem 6 shows that the estimator is consistent over the randomization distribution, and asymptotically normal as the population size grows larger.

\begin{theorem}
     Suppose that potential outcomes are bounded. Under Assumptions 6-10,

\begin{equation*}
    \frac{\sqrt{N}\left \{ \widehat{\overline{\tau}}_{C_{t-p:t},t}(\textbf{d},\tilde{\textbf{d}};p)-\overline{\tau}_{C_{t-p:},t}(\textbf{d},\tilde{\textbf{d}};p) \right \}}{\sigma_{t}(\textbf{d},\tilde{\textbf{d}};p)}\overset{d}{\rightarrow} \mathcal{N}(0,1),\ as\ N\rightarrow\infty
\end{equation*}

where $\sigma_{t}(\textbf{d},\tilde{\textbf{d}};p)$ is defined in Section 6 of Appendix A.

\end{theorem}

Once again, I use \cite{bloom} confidence intervals for hypothesis testing. In Section 4.2, I conduct Monte Carlo simulations to study the asymptotic properties of these estimators.

\section{Monte Carlo Simulations}

In this section I show the desirable finite-sample properties of the proposed nonparametric estimators for the factors $k-p$ to $k$ causal effects and the local lag-$p$ dynamic causal effects.  I study their finite-sample properties in terms of the average bias (Av. Bias), median bias (Med. Bias), root mean-squared error (RMSE), coverage of the Confidence Interval (Cover) and the Confidence Interval length (CIL).

\subsection{Factorial Designs}

I consider a setting with $N=1.000$ units and $K=2$ binary factors. For each factor, assignment is randomized following a Bernoulli distribution:

\begin{equation*}
    \mathbb{P}\left ( Z_{i,k}=z_{k}\right )\propto \prod_{i\in[N]}p_{i,k}^{z_{i,k}}(1-p_{i,k})^{1-z_{i,k}} 
\end{equation*}

I set $p_{i,1}=p_{1}=0.5$ and $p_{i,2}=p_{2}=0.5$, such that the data generating process emulates a completely randomized experiment. Potential outcomes are generated according to the following linear model:

\begin{equation*}
    Y_{i}(d_{1},d_{2})=\beta_{0}+\beta_{1}d_{1}+\beta_{2}d_{2}+\beta_{1,2}d_{1}d_{2}+\varepsilon_{i}
\end{equation*}

I set $\beta_{0}=0, \beta_{1}=0.5,\beta_{2}=1,\beta_{1,2}=0.25$ and $\varepsilon_{i}\sim \mathcal{N}(0,1)$. Potential treatments $D_{i,k}(z_{k})$ are generated following Bernoulli distributions with parameters $\delta_{k}(z_{k})$. For $k=1$ I set $\delta_1(z_{1})=0.2+0.7z_{1}$, such that compliance rate is 70\%. For $k=1$ I set $\delta_2(z_{2})=0.2+0.6z_{2}$, such that compliance rate is 60\%. I then conduct 10.000 Monte Carlo experiments to evaluate the performance of the Horvitz-Thompson estimators proposed in Section 2.3, the results are summarized in Table 1.

\begin{table}[t]
\caption{Simulation results for Causal Effects Estimators in Factorial Designs}
\centering
\begin{tabular}{c|ccccc}
\hline
Causal Effect & $\tau_{C_{1:2}}(\left\{1,1\right\}\left\{0,0\right\})$ & $\tau_{C_{1:2}}(\left\{1,0\right\}\left\{0,0\right\})$ & $\tau_{C_{1:2}}(\left\{0,1\right\}\left\{0,0\right\})$ & $\tau_{C_{1}}(1,0)$   & $\tau_{C_{2}}(1,0)$  \\ \hline
Av. Bias      & -0.001     & 0.003      & 0.002      & 0.002  & 0.005 \\
Med. Bias     & 0.001      & 0.001      & 0.003      & -0.010 & 0.012 \\
RMSE          & 0.178      & 0.187      & 0.169      & 0.141  & 0.142 \\
Cover         & 0.941      & 0.931      & 0.948      & 0.938  & 0.956 \\
CIL           & 0.824      & 0.795      & 0.801      & 0.359  & 0.422 \\ \hline
\end{tabular}
\\
\scriptsize \noindent \textit{Note:} Simulations based on 10.000 Monte Carlo experiments with sample size $N=1.000$ and $K=2$.
\end{table}

The first three columns present the results for the estimators of factors 1 to 2 causal effects associated to three different treatment uptakes using the multiple-differences approach for response functions separately and then generating the estimate of the causal effect by taking the difference across response functions. Note that in the case of $K=2$, the multiple-differences estimand takes the form of a two-stage difference-in-differences across assignments.

The estimator exhibits negligible bias for the three causal effects and Bloom confidence interval have empirical coverage close to the desired 95\% coverage, with lengths fairly stable across the different effects.

The last two columns present the results for the two-stage Horvitz-Thompson estimator of the factor $k$ specific causal effects. Again, the estimator exhibits negligible bias across the simulation and Bloom confidence interval have a good performance in terms of coverage. Confidence interval have approximately have the length of the confidence intervals from the multiple-differences estimator.

\subsection{Panel Experiments}

I consider a balanced panel setting with $N=1000$ and $T=2$. The simulation focuses on the lag-p dynamic causal effects with $p=t$, that is, the dynamic causal effects associated to the whole treatment path in the setting. Assignment is sequentially randomized following a Bernoulli distribution:

\begin{equation*}
    \mathbb{P}\left ( Z_{i,t}=z_{t}|Z_{i,1:t-1}=z_{1:t-1},Y_{i,1:t-1}=y_{1:t-1},D_{i,1:t-1}=d\right )\propto \prod_{i\in[N]}p_{i,t}^{z_{i,t}}(1-p_{i,t})^{1-z_{i,t}} 
\end{equation*}

I set $p_{i,t}=p_{i}=0.6$. For the choice model. Outcomes in period $t$ are specified to have the following linear working model:

\begin{equation*}
    Y_{i,t}=\delta_{t}'Y_{i,1:t-1}+\beta_{1:t}'D_{i,1:t}+U_{i,t}(D_{i,1:t})
\end{equation*}

I set $\beta_{t}=1$ and $\beta_{t-1}=0.5$, the vector $\beta_{1:t-2}$ to be a vector of zeros, and $U_{i,t}(D_{1:t})\sim \mathcal{N}(0,1)$. Potential treatments at  period $t$ are generated in a way that the compliance rate is 50\% at each time period.

In Table 2, I compare the performance of the proposed Horvitz-Thompson estimator (HT) with the performance of the period-specific 2SLS estimator\footnote{In Section C of the Appendix I present a causal decomposition of the period-specific 2SLS estimand and show that it does not hold a straightforward causal interpretation in $t=2$.}, taking the local lag-0 dynamic causal effect as the target parameter.

\begin{table}[t]
\caption{Simulation results for the Lag-0 dynamic causal effect}
\centering
\begin{tabular}{c|cc|cc|cc}
\hline
Time-$t$ effect & \multicolumn{2}{c|}{$t=1$} & \multicolumn{2}{c|}{$t=2$} & \multicolumn{2}{c}{Total} \\ \hline
Estimator     & HT          & 2SLS       & HT          & 2SLS       & HT          & 2SLS        \\ \hline
Av. Bias      & -0.003      & 0.003      & 0.027       & 0.368      & 0.015       & 0.156       \\
Med. Bias     & -0.002      & 0.003      & 0.034       & 0.392      & 0.019       & 0.163       \\
RMSE          & 0.097       & 0.081      & 0.119       & 0.484      & 0.237       & 0.169       \\
Cover         & 0.944            & 0.948      &     0.952        & 0.612      &      0.955       & 0.784       \\
CIL           &   0.356          & 0.223      &     0.508        & 0.285      &   0.441          & 0.258       \\ \hline
\end{tabular}
\\
\scriptsize \noindent \textit{Note:} Simulations based on 10.000 Monte Carlo experiments with sample size $N=1.000$ and $T=2$. CIs for the 2SLS estimator were built using the standard estimated variance using the Delta Method.
\end{table}

The first two columns show the results for the first period. When $t=1$, dynamics play no role in the model. Hence, both the nonparametric estimator and the static 2SLS show little to none Monte Carlo bias. Moreover, the coverage is close to the desired 95\%, with the 2SLS estimator showing a tighter Confidence Interval on average. When it comes to the second period, the Horvitz-Thompson estimator remains consistent in the Monte Carlo exercise, as shown by the third column. The period-specific 2SLS estimator is severely biased, with coverage far from the desired 95\%. The last two columns stack the lag-0 estimates across the two time periods. Thus, the results can be interpreted as a weighted average of the time-$t$ results, which explains why the performance of the 2SLS estimator is better than in the second period alone. As the number of periods grows larger, however, one should expect the 2SLS estimator to perform increasingly worse with the number of time periods.

In Table 4, I present the simulation results for different time-$t$ local lag-1 dynamic causal effects. The lag-1 effects can only be estimated for period 2. I present results for the difference-in-differences modified Horvitz-Thompson estimator (HT) and a multivariate 2SLS estimator (MV2SLS). I consider the performance of the estimators with respect to effects of full exposure in the first two columns, exposure in the second period in the third and fourth columns, and exposure in the first period in the last two columns.

The multivariate 2SLS specification yields substantially biased estimates for the three different target parameters. Coverage is closer to the desired 95\% than in the simulations for the period-specific 2SLS. However, it is never greater than 81.5\%.

The proposed nonparametric estimator exhibits great Monte Carlo performance. The average bias, median bias and root mean-squared error for the causal effects are small, and the coverage of the conservative confidence interval is close to the desired 95\% coverage. Confidence interval lengths are fairly stable across the considered treatment paths.

\begin{table}[t]
\caption{Simulation results for the Lag-1 dynamic causal effects}
\centering
\begin{tabular}{c|cccccc}
\hline
Lag-1 effect & \multicolumn{2}{c}{$\tau_{C_{1:2}}(\left\{1,1\right\},\left\{0,0\right\})$} & \multicolumn{2}{c}{$\tau_{C_{1:2}}(\left\{0,1\right\},\left\{0,0\right\})$} & \multicolumn{2}{c}{$\tau_{C_{1:2}}(\left\{1,0\right\},\left\{0,0\right\})$} \\ \hline
Estimator & HT      & MV2SLS   & HT      & MV2SLS   & HT      & MV2SLS   \\ \hline
Av. Bias  & -0.0008 & 0.1968 & -0.0039 & 0.1306 & -0.0051 & 0.1641 \\
Med. Bias & -0.0012 & 0.1970 & 0.0002  & 0.1312 & -0.0046 & 0.1646 \\
RMSE      & 0.0772  & 0.1975 & 0.0762  & 0.1404 & 0.0845  & 0.1937 \\
Cover     & 0.938   & 0.813  & 0.956   & 0.798  & 0.928   & 0.815  \\
CIL       & 0.8721  & 0.3019 & 0.8023  & 0.3145 & 0.8674  & 0.3222 \\ \hline
\end{tabular}
\\
\scriptsize \noindent \textit{Note:} Simulations based on 10.000 Monte Carlo experiments with sample size $N=1.000$ and $T=2$. CIs for the 2SLS estimator were built using the standard estimated variance using the Delta Method.
\end{table}

Overall, the Monte Carlo Simulations assert the desirable finite-sample performance of the proposed estimators for dynamic causal effects over the randomization, while bringing evidence of pitfalls associated to the standard 2SLS methods in the presence of time-varying heterogeneity.

\section{Application - GOTV Experiment}

In this section, I use the estimators introduced in Section 2 to revisit the causal effects of get-out-the-vote (GOTV) efforts on youth voting turnout in the large field experiment conducted through the Youth Vote Coalition by \cite{nick}.

In this experiment individual registered voters were randomly assigned to receive a call from a volunteer phone bank ($Z_{i,1}$), a professional phone bank ($Z_{i,2}$), both or neither. The probabilities of assignment for $Z_{i,1}$ and $Z_{i,2}$ are 0.5 each, which means the four possible paths of assignment are observed with the same probability. The outcome of interest is turnout in the 2004 presidential election. Imperfect compliance in this setting arises from the fact that not all individuals actually picked up the phone when received the call.

In order to estimate the causal effects of the phone call encouragements on turnout, I follow \cite{Blackwell2017} and focus on college-age respondents from sites in which the experiment was implemented successfully (both forms of contact were assigned and there are no violations of the exclusion restriction), leading to $N=26,974$ respondents. Table 4 summarizes the results.

\begin{table}[t]
\caption{Causal Effects of GOTV Efforts on Voting Probability}
\centering
\begin{tabular}{c|cc}
\hline
Causal Effect & Estimate & 95\% CI          \\ \hline
$\tau_{C_{1:2}}(\left\{1,1\right\},\left\{0,0\right\})$    & 0.113    & (0.025, 0.201)  \\
$\tau_{C_{1:2}}(\left\{1,0\right\},\left\{0,0\right\})$   & 0.136    & (0.042, 0.229)  \\
$\tau_{C_{1:2}}(\left\{0,1\right\},\left\{0,0\right\})$    & 0.117    & (0.026, 0.208)  \\
$\tau_{C_{1}}(1,0)$          & 0.038    & (0.016, 0.059)  \\
$\tau_{C_{2}}(1,0)$          & 0.018    & (-0.007, 0.043) \\ \hline
\end{tabular}\\
\scriptsize \noindent \textit{Note:} Estimates for the causal effects from the Nickerson (2007) youth GOTV experiment. $N = 26,974$ registered voters aged 17-22. 
\end{table}

The first three rows of Table 4 show that volunteer and professional phone bank calls had a positive impact on turnout. The causal effect of receiving both calls vs not receiving any call is similar in magnitude to the causal effect of receiving only one type of calls vs no calls, which suggests that these two interventions are not complementary. The last two rows present the factor $k$ specific causal effects, which are far smaller in magnitude. The fact that the estimates for $\tau_{C_{1}}(1,0)$ and $\tau_{C_{2}}(1,0)$  are smaller than the estimates for $\tau_{C_{1:2}}(\left\{1,0\right\},\left\{0,0\right\})$ and $\tau_{C_{1:2}}(\left\{0,1\right\},\left\{0,0\right\})$ suggest that there are negative effect of the interaction of treatments, indicating diminishing returns to GOTV efforts, which has been previously documented by \cite{Blackwell2017}.

With the exception of $\tau_{C_{2}}(1,0)$, all the estimates for causal effects are statistically significant. I use Bloom confidence intervals for hypothesis testing. The joint compliance rates are estimated to be around 22.5\%, which indicates that Bloom CIs should exhibit coverage rate close to the desired 95\%. The estimates for the joint compliance rate are stable across the modified first stages, as displayed in Table 5. When $K=2$, the modified first-stage has expectation $\pm\frac{1}{N}\left | C_{1:2}\right |$, where the negative signs in the two middle rows arise mechanically from the multiple-differences operator, as shown in Section 1 of Appendix A. Thus, the estimates for compliance rates should be interpreted as the absolute values displayed in Table 5.

\begin{table}[t]
\caption{Estimates of Joint Compliance Rates}
\centering
\begin{tabular}{c|cc}
\hline
Modified First Stage & Estimate & 95\% CI           \\ \hline
$m_{1:2}^{FS}(1,1)$               & 0.230    & (0.218, 0.241)   \\
$m_{1:2}^{FS}(1,0)$               & -0.241   & (-0.257, -0.225) \\
$m_{1:2}^{FS}(0,1)$               & -0.228   & (-0.240, -0.215) \\
$m_{1:2}^{FS}(0,0)$               & 0.203    & (0.182, 0.225)   \\ \hline
\end{tabular}\\
\scriptsize \noindent \textit{Note:} Modified first-stage estimates of joint compliers  from the Nickerson (2007) youth GOTV experiment. $N = 26,974$ registered voters aged 17-22. 
\end{table}

Overall, the result of the application align to the previous results regarding the 2004 Youth Vote Coalition GOTV experiment and show evidence of diminishing returns to follow-up contact on turnout. The fundamental difference between the previous results and the results presented in this Section comes from the approach towards uncertainty. While \cite{nick} and \cite{Blackwell2017} construct confidence intervals based on a traditional sampling perspective, I think of the population of the experiment as fixed, and the phone call assignments as stochastic. The framework from Section 2 implies that if Assumptions 1-5 hold for this finite population, then the CIs from Table 4 be interpreted as a valid, but possibly conservative 95\% confidence intervals for the causal effects of interest.

\section{Conclusion}

This paper develops a design-based framework for experimental settings with imperfect compliance in which multiple treatments can affect the outcome of interest. I focus on the cases of factorial designs and panel experiments and show that under standard instrumental variable assumptions and a treatment exclusion restriction which can be justified in both settings, causal effects associated to sequences of treatments can be identified by exploiting variation in the sequence of assignments associated to it in what I define as a two-stage multiple-differences estimand.

I introduce nonparametric estimators for causal effects which are consistent over the randomization distribution, and derive their  finite population asymptotic distributions. I illustrate the desirable properties of the proposed estimators through Monte Carlo simulation studies and apply the estimator for factorial designs in a famous political science setting. The results show that the causal effects of GOTV efforts are large and statistically significant under potentially conservative inference procedures.

\bibliography{bibliography.bib}

\section*{Appendix A - Main Proofs}

\subsection*{Proof of Theorem 1}

For the first part, note that for any $k\in[K]$, under Assumptions 1-5,

\begin{align*}
    &\mathbb{E}\left [ \frac{1}{N}\sum_{i=1}^{N}\frac{Z_{i,k}D_{i,k}}{\pi_{i,k}(1)}-\frac{1}{N}\sum_{i=1}^{N}\frac{(1-Z_{i,k})D_{i,k}}{\pi_{i,k}(0)}|\mathcal{F}_{1:N,-k} \right ]\\&=\frac{1}{N}\sum_{i=1}^{N}D_{i,k}(1)-\frac{1}{N}\sum_{i=1}^{N}D_{i,k}(0)\\&=\frac{1}{N}\left | C_{k}\right |
\end{align*}

and 
\begin{align*}
    &\mathbb{E}\left [ \frac{1}{N}\sum_{i=1}^{N}\frac{Z_{i,k}Y_{i}}{\pi_{i,k}(1)}-\frac{1}{N}\sum_{i=1}^{N}\frac{(1-Z_{i,k})Y_{i}}{\pi_{i,k}(0)}|\mathcal{F}_{1:N,-k} \right ]\\&=\frac{1}{N}\sum_{i=1}^{N}Y_{i}(d^{0bs}_{i,-k},D_{i,k}(1))-\frac{1}{N}\sum_{i=1}^{N}Y_{i}(d^{obs}_{i,-k},D_{i,k}(0))\\&=\frac{1}{N}\sum_{i\in C_{k}}Y_{i}(d^{obs}_{i,-k},1)-Y_{i}(d^{obs}_{i,-k},0)
\end{align*}

For the second part, I conduct the proof by induction. First, I show that all potential outcomes of interested are identified in the case of $p=1$\footnote{I omit the conditioning on the filtration for the sake of simplicity in notation, but all expectations are taken conditional on $\mathcal{F}_{1:N,-(k-1:k)}$}. Then, I show that if the result holds for a general $p$, then it must hold for $p+1$.

I begin with the case where $(D_{i,k-1},D_{i,k})=(1,1)$. The first stage is

{\footnotesize\begin{align*}
    &\mathbb{E}\left [ \frac{1}{N}\sum_{i=1}^{N}\frac{\mathbf{1}\left\{ Z_{i,k-1:k}=(1,1)\right\}\mathbf{1}\left\{D_{i,k-1:k}=(1,1) \right\}}{\pi_{i,k-1:k}(1,1)} \right ]-\mathbb{E}\left [ \frac{1}{N}\sum_{i=1}^{N}\frac{\mathbf{1}\left\{ Z_{i,k-1:k}=(1,0)\right\}\mathbf{1}\left\{D_{i,k-1:k}=(1,1) \right\}}{\pi_{i,k-1:k}(1,0)} \right ]\\&-\left(\mathbb{E}\left [ \frac{1}{N}\sum_{i=1}^{N}\frac{\mathbf{1}\left\{ Z_{i,k-1:k}=(0,1)\right\}\mathbf{1}\left\{D_{i,k-1:k}=(1,1) \right\}}{\pi_{i,k-1:k}(0,1)} \right ]-\mathbb{E}\left [ \frac{1}{N}\sum_{i=1}^{N}\frac{\mathbf{1}\left\{ Z_{i,k-1:k}=(0,0)\right\}\mathbf{1}\left\{D_{i,k-1:k}=(1,1) \right\}}{\pi_{i,k-1:k}(0,0)} \right ]\right)\\&=\frac{1}{N}\sum_{i=1}^{N}D_{i,k-1}(1)D_{i,k}(1)-\frac{1}{N}\sum_{i=1}^{N}D_{i,k-1}(1)D_{i,k}(0)-\left(\frac{1}{N}\sum_{i=1}^{N}D_{i,k-1}(0)D_{i,k}(1)-\frac{1}{N}\sum_{i=1}^{N}D_{i,k-1}(0)D_{i,k}(0)\right)\\&=\frac{1}{N}\sum_{i=1}^{N}\mathbf{1}\left\{D_{i,k-1}(1)=1,i\in C_{k} \right\}-\frac{1}{N}\sum_{i=1}^{N}\mathbf{1}\left\{D_{i,k-1}(0)=1,i\in C_{k} \right\}\\&=\frac{1}{N}\sum_{i=1}^{N}\mathbf{1}\left\{i\in C_{k-1:k} \right\}=\frac{1}{N}\left |C_{k-1:k} \right |
\end{align*}}

where the first equality follows from Assumptions 2 (ii) and 4, the second and third equality follow from Assumption 3.

For the reduced form, note that Assumptions 1 and 2 (i) combined with the assumptions above imply that

{\footnotesize \begin{align*}
    & \mathbb{E}\left [ \frac{1}{N}\sum_{i=1}^{N}\frac{Y_{i}\mathbf{1}\left\{ Z_{i,k-1:k}=(1,1)\right\}\mathbf{1}\left\{D_{i,k-1:k}=(1,1) \right\}}{\pi_{i,k-1:k}(1,1)} \right ]-\mathbb{E}\left [ \frac{1}{N}\sum_{i=1}^{N}\frac{Y_{i}\mathbf{1}\left\{ Z_{i,k-1:k}=(1,0)\right\}\mathbf{1}\left\{D_{i,k-1:k}=(1,1) \right\}}{\pi_{i,k-1:k}(1,0)} \right ]\\&-\left(\mathbb{E}\left [ \frac{1}{N}\sum_{i=1}^{N}\frac{Y_{i}\mathbf{1}\left\{ Z_{i,k-1:k}=(0,1)\right\}\mathbf{1}\left\{D_{i,k-1:k}=(1,1) \right\}}{\pi_{i,k-1:k}(0,1)} \right ]-\mathbb{E}\left [ \frac{1}{N}\sum_{i=1}^{N}\frac{Y_{i}\mathbf{1}\left\{ Z_{i,k-1:k}=(0,0)\right\}\mathbf{1}\left\{D_{i,k-1:k}=(1,1) \right\}}{\pi_{i,k-1:k}(0,0)} \right ]\right)\\&=\frac{1}{N}\sum_{i\in C_{k-1:k}}Y_{i}(d^{obs}_{i,-(k-1:k)},1,1)
\end{align*}}

and therefore, the ratio identifies

\begin{equation*}
    \frac{1}{\left |C_{k-1:k} \right |}\sum_{i\in C_{k-1:k}}Y_{i}(d^{obs}_{i,-(k-1:k)},1,1)=m_{k-1:k}(1,1)
\end{equation*}

For the case where $(D_{i,k-1},D_{i,k})=(1,0)$, using the same reasoning, we find that the first stage is

{\footnotesize\begin{align*}
    &\mathbb{E}\left [ \frac{1}{N}\sum_{i=1}^{N}\frac{\mathbf{1}\left\{ Z_{i,k-1:k}=(1,1)\right\}\mathbf{1}\left\{D_{i,k-1:k}=(1,0) \right\}}{\pi_{i,k-1:k}(1,1)} \right ]-\mathbb{E}\left [ \frac{1}{N}\sum_{i=1}^{N}\frac{\mathbf{1}\left\{ Z_{i,k-1:k}=(1,0)\right\}\mathbf{1}\left\{D_{i,k-1:k}=(1,0) \right\}}{\pi_{i,k-1:k}(1,0)} \right ]\\&-\left(\mathbb{E}\left [ \frac{1}{N}\sum_{i=1}^{N}\frac{\mathbf{1}\left\{ Z_{i,k-1:k}=(0,1)\right\}\mathbf{1}\left\{D_{i,k-1:k}=(1,0) \right\}}{\pi_{i,k-1:k}(0,1)} \right ]-\mathbb{E}\left [ \frac{1}{N}\sum_{i=1}^{N}\frac{\mathbf{1}\left\{ Z_{i,k-1:k}=(0,0)\right\}\mathbf{1}\left\{D_{i,k-1:k}=(1,0) \right\}}{\pi_{i,k-1:k}(0,0)} \right ]\right)\\&=\frac{1}{N}\sum_{i=1}^{N}D_{i,k-1}(1)(1-D_{i,k}(1))-\frac{1}{N}\sum_{i=1}^{N}D_{i,k-1}(1)(1-D_{i,k}(0))\\&-\left(\frac{1}{N}\sum_{i=1}^{N}D_{i,k-1}(0)(1-D_{i,k}(1))-\frac{1}{N}\sum_{i=1}^{N}D_{i,k-1}(0)(1-D_{i,k}(0))\right)\\&=-\frac{1}{N}\sum_{i=1}^{N}\mathbf{1}\left\{D_{i,k-1}(1)=1,i\in C_{k} \right\}+\frac{1}{N}\sum_{i=1}^{N}\mathbf{1}\left\{D_{i,k-1}(0)=1,i\in C_{k} \right\}\\&=-\frac{1}{N}\sum_{i=1}^{N}\mathbf{1}\left\{i\in C_{k-1:k} \right\}=-\frac{1}{N}\left |C_{k-1:k} \right |
\end{align*}}

Similarly, the reduced form identifies

{\footnotesize \begin{align*}
    & \mathbb{E}\left [ \frac{1}{N}\sum_{i=1}^{N}\frac{Y_{i}\mathbf{1}\left\{ Z_{i,k-1:k}=(1,1)\right\}\mathbf{1}\left\{D_{i,k-1:k}=(1,0) \right\}}{\pi_{i,k-1:k}(1,1)} \right ]-\mathbb{E}\left [ \frac{1}{N}\sum_{i=1}^{N}\frac{Y_{i}\mathbf{1}\left\{ Z_{i,k-1:k}=(1,0)\right\}\mathbf{1}\left\{D_{i,k-1:k}=(1,0) \right\}}{\pi_{i,k-1:k}(1,0)} \right ]\\&-\left(\mathbb{E}\left [ \frac{1}{N}\sum_{i=1}^{N}\frac{Y_{i}\mathbf{1}\left\{ Z_{i,k-1:k}=(0,1)\right\}\mathbf{1}\left\{D_{i,k-1:k}=(1,0) \right\}}{\pi_{i,k-1:k}(0,1)} \right ]-\mathbb{E}\left [ \frac{1}{N}\sum_{i=1}^{N}\frac{Y_{i}\mathbf{1}\left\{ Z_{i,k-1:k}=(0,0)\right\}\mathbf{1}\left\{D_{i,k-1:k}=(1,0) \right\}}{\pi_{i,k-1:k}(0,0)} \right ]\right)\\&=-\frac{1}{N}\sum_{i\in C_{k-1:k}}Y_{i}(d^{obs}_{i,-(k-1:k)},1,0)
\end{align*}}

and therefore, the ratio identifies

\begin{equation*}
    \frac{1}{\left |C_{k-1:k} \right |}\sum_{i\in C_{k-1:k}}Y_{i}(d^{obs}_{i,-(k-1:k)},1,0)=m_{k-1:k}(1,0)
\end{equation*}

For the case of $(D_{i,k-1},D_{i,k})=(0,1)$, the first stage identifies

{\footnotesize\begin{align*}
    &\mathbb{E}\left [ \frac{1}{N}\sum_{i=1}^{N}\frac{\mathbf{1}\left\{ Z_{i,k-1:k}=(1,1)\right\}\mathbf{1}\left\{D_{i,k-1:k}=(0,1) \right\}}{\pi_{i,k-1:k}(1,1)} \right ]-\mathbb{E}\left [ \frac{1}{N}\sum_{i=1}^{N}\frac{\mathbf{1}\left\{ Z_{i,k-1:k}=(1,0)\right\}\mathbf{1}\left\{D_{i,k-1:k}=(0,1) \right\}}{\pi_{i,k-1:k}(1,0)} \right ]\\&-\left(\mathbb{E}\left [ \frac{1}{N}\sum_{i=1}^{N}\frac{\mathbf{1}\left\{ Z_{i,k-1:k}=(0,1)\right\}\mathbf{1}\left\{D_{i,k-1:k}=(0,1) \right\}}{\pi_{i,k-1:k}(0,1)} \right ]-\mathbb{E}\left [ \frac{1}{N}\sum_{i=1}^{N}\frac{\mathbf{1}\left\{ Z_{i,k-1:k}=(0,0)\right\}\mathbf{1}\left\{D_{i,k-1:k}=(0,1) \right\}}{\pi_{i,k-1:k}(0,0)} \right ]\right)\\&=\frac{1}{N}\sum_{i=1}^{N}(1-D_{i,k-1})(1)D_{i,k}(1)-\frac{1}{N}\sum_{i=1}^{N}(1-D_{i,k-1}(1))D_{i,k}(0)\\&-\left(\frac{1}{N}\sum_{i=1}^{N}(1-D_{i,k-1}(0))D_{i,k}(1)-\frac{1}{N}\sum_{i=1}^{N}(1-D_{i,k-1}(0))D_{i,k}(0)\right)\\&=\frac{1}{N}\sum_{i=1}^{N}\mathbf{1}\left\{D_{i,k-1}(1)=0,i\in C_{k} \right\}-\frac{1}{N}\sum_{i=1}^{N}\mathbf{1}\left\{D_{i,k-1}(0)=0,i\in C_{k} \right\}\\&=-\frac{1}{N}\sum_{i=1}^{N}\mathbf{1}\left\{i\in C_{k-1:k} \right\}=-\frac{1}{N}\left |C_{k-1:k} \right |
\end{align*}}

The reduced form identifies

{\footnotesize \begin{align*}
    & \mathbb{E}\left [ \frac{1}{N}\sum_{i=1}^{N}\frac{Y_{i}\mathbf{1}\left\{ Z_{i,k-1:k}=(1,1)\right\}\mathbf{1}\left\{D_{i,k-1:k}=(0,1) \right\}}{\pi_{i,k-1:k}(1,1)} \right ]-\mathbb{E}\left [ \frac{1}{N}\sum_{i=1}^{N}\frac{Y_{i}\mathbf{1}\left\{ Z_{i,k-1:k}=(1,0)\right\}\mathbf{1}\left\{D_{i,k-1:k}=(0,1) \right\}}{\pi_{i,k-1:k}(1,0)} \right ]\\&-\left(\mathbb{E}\left [ \frac{1}{N}\sum_{i=1}^{N}\frac{Y_{i}\mathbf{1}\left\{ Z_{i,k-1:k}=(0,1)\right\}\mathbf{1}\left\{D_{i,k-1:k}=(0,1) \right\}}{\pi_{i,k-1:k}(0,1)} \right ]-\mathbb{E}\left [ \frac{1}{N}\sum_{i=1}^{N}\frac{Y_{i}\mathbf{1}\left\{ Z_{i,k-1:k}=(0,0)\right\}\mathbf{1}\left\{D_{i,k-1:k}=(0,1) \right\}}{\pi_{i,k-1:k}(0,0)} \right ]\right)\\&=-\frac{1}{N}\sum_{i\in C_{k-1:k}}Y_{i}(d^{obs}_{i,-(k-1:k)},0,1)
\end{align*}}

and therefore, the ratio identifies

\begin{equation*}
    \frac{1}{\left |C_{k-1:k} \right |}\sum_{i\in C_{k-1:k}}Y_{i}(d^{obs}_{i,-(k-1:k)},0,1)=m_{k-1:k}(0,1)
\end{equation*}

Finally, for the case of $(D_{i,k-1},D_{i,k})=(0,0)$, the first stage identifies

{\footnotesize\begin{align*}
    &\mathbb{E}\left [ \frac{1}{N}\sum_{i=1}^{N}\frac{\mathbf{1}\left\{ Z_{i,k-1:k}=(1,1)\right\}\mathbf{1}\left\{D_{i,k-1:k}=(0,0) \right\}}{\pi_{i,k-1:k}(1,1)} \right ]-\mathbb{E}\left [ \frac{1}{N}\sum_{i=1}^{N}\frac{\mathbf{1}\left\{ Z_{i,k-1:k}=(1,0)\right\}\mathbf{1}\left\{D_{i,k-1:k}=(0,0) \right\}}{\pi_{i,k-1:k}(1,0)} \right ]\\&-\left(\mathbb{E}\left [ \frac{1}{N}\sum_{i=1}^{N}\frac{\mathbf{1}\left\{ Z_{i,k-1:k}=(0,1)\right\}\mathbf{1}\left\{D_{i,k-1:k}=(0,0) \right\}}{\pi_{i,k-1:k}(0,1)} \right ]-\mathbb{E}\left [ \frac{1}{N}\sum_{i=1}^{N}\frac{\mathbf{1}\left\{ Z_{i,k-1:k}=(0,0)\right\}\mathbf{1}\left\{D_{i,k-1:k}=(0,0) \right\}}{\pi_{i,k-1:k}(0,0)} \right ]\right)\\&=\frac{1}{N}\sum_{i=1}^{N}(1-D_{i,k-1})(1)(1-D_{i,k}(1))-\frac{1}{N}\sum_{i=1}^{N}(1-D_{i,k-1}(1))(1-D_{i,k}(0))\\&-\left(\frac{1}{N}\sum_{i=1}^{N}(1-D_{i,k-1}(0))(1-D_{i,k}(1))-\frac{1}{N}\sum_{i=1}^{N}(1-D_{i,k-1}(0))(1-D_{i,k}(0))\right)\\&=-\frac{1}{N}\sum_{i=1}^{N}\mathbf{1}\left\{D_{i,k-1}(1)=0,i\in C_{k} \right\}+\frac{1}{N}\sum_{i=1}^{N}\mathbf{1}\left\{D_{i,k-1}(0)=0,i\in C_{k} \right\}\\&=\frac{1}{N}\sum_{i=1}^{N}\mathbf{1}\left\{i\in C_{k-1:k} \right\}=\frac{1}{N}\left |C_{k-1:k} \right |
\end{align*}}

The reduced form identifies

{\footnotesize \begin{align*}
    & \mathbb{E}\left [ \frac{1}{N}\sum_{i=1}^{N}\frac{Y_{i}\mathbf{1}\left\{ Z_{i,k-1:k}=(1,1)\right\}\mathbf{1}\left\{D_{i,k-1:k}=(0,0) \right\}}{\pi_{i,k-1:k}(1,1)} \right ]-\mathbb{E}\left [ \frac{1}{N}\sum_{i=1}^{N}\frac{Y_{i}\mathbf{1}\left\{ Z_{i,k-1:k}=(1,0)\right\}\mathbf{1}\left\{D_{i,k-1:k}=(0,0) \right\}}{\pi_{i,k-1:k}(1,0)} \right ]\\&-\left(\mathbb{E}\left [ \frac{1}{N}\sum_{i=1}^{N}\frac{Y_{i}\mathbf{1}\left\{ Z_{i,k-1:k}=(0,1)\right\}\mathbf{1}\left\{D_{i,k-1:k}=(0,0) \right\}}{\pi_{i,k-1:k}(0,1)} \right ]-\mathbb{E}\left [ \frac{1}{N}\sum_{i=1}^{N}\frac{Y_{i}\mathbf{1}\left\{ Z_{i,k-1:k}=(0,0)\right\}\mathbf{1}\left\{D_{i,k-1:k}=(0,0) \right\}}{\pi_{i,k-1:k}(0,0)} \right ]\right)\\&=\frac{1}{N}\sum_{i\in C_{k-1:k}}Y_{i}(d^{obs}_{i,-(k-1:k)},0,0)
\end{align*}}

and therefore, the ratio identifies

\begin{equation*}
    \frac{1}{\left |C_{k-1:k} \right |}\sum_{i\in C_{k-1:k}}Y_{i}(d^{obs}_{i,-(k-1:k)},0,0)=m_{k-1:k}(0,0)
\end{equation*}

which concludes the proof for the case of $p=1$. For a general $p+1$, I prove the result only for $D_{i,k-p-1}=1$. The case for $D_{i,k-p-1}=0$ is analogous  and can be derived using the intuition from the displayed result above. Let $s(\textbf{d})\in \left\{-1,1\right\}$ be a sign factor determined by the pattern of zeros and ones in $\mathbf{d}$. I begin with the first stage: 

\begin{align*}
    &\Delta^{p+2}_{HT}\left ( \mathbb{E}\left [ \frac{1}{N}\sum_{i=1}^{N}\frac{\mathbf{1}\left\{ Z_{i,k-p-1:k}=\textbf{z}\right\}\mathbf{1}\left\{ D_{i,k-p-1:k}=(1,\textbf{d})\right\}}{\pi_{i,k-p-1:k}(\textbf{z})}|\mathcal{F}_{1:N,-(k-p-2)} \right ] \right )\\&=s(\textbf{d})\Delta^{p+1}_{HT}\left ( \mathbb{E}\left [ \frac{1}{N}\sum_{i=1}^{N}\frac{\mathbf{1}\left\{ Z_{i,k-p-1:k}=(1,\textbf{z})\right\}\mathbf{1}\left\{ D_{i,k-p-1:k}=(1,\textbf{d})\right\}}{\pi_{i,k-p-1:k}(1,\textbf{z})}|\mathcal{F}_{1:N,-(k-p-2)} \right ] \right )\\&-s(\textbf{d})\Delta^{p+1}_{HT}\left ( \mathbb{E}\left [ \frac{1}{N}\sum_{i=1}^{N}\frac{\mathbf{1}\left\{ Z_{i,k-p-1:k}=(0,\textbf{z})\right\}\mathbf{1}\left\{ D_{i,k-p-1:k}=(1,\textbf{d})\right\}}{\pi_{i,k-p-1:k}(0,\textbf{z})}|\mathcal{F}_{1:N,-(k-p-2)} \right ] \right )\\&=s(\textbf{d})\frac{1}{N}\sum_{i=1}^{N}D_{i,k-p-1}(1)\mathbf{1}\left\{ i\in C_{k-p:k}\right\}-s(\textbf{d})\frac{1}{N}\sum_{i=1}^{N}D_{i,k-p-1}(0)\mathbf{1}\left\{ i\in C_{k-p:k}\right\}\\&=s(\textbf{d})\frac{1}{N}\sum_{i=1}^{N}\mathbf{1}\left\{ i\in C_{k-p-1:k}\right\}=s(\textbf{d})\frac{1}{N}\left | C_{k-p-1:k}\right |
\end{align*}

\noindent where the first equality follows from the definition of $\Delta^{p+2}(.)$, the second equality follows from Assumptions 2 (ii), 4 and assuming the result holds for $p$ and the third equality follows from Assumption 3.

For the modified reduced form, note that 

\begin{align*}
    &\Delta^{p+2}_{HT}\left ( \mathbb{E}\left [ \frac{1}{N}\sum_{i=1}^{N}\frac{\mathbf{1}\left\{ Z_{i,k-p-1:k}=\textbf{z}\right\}\mathbf{1}\left\{ D_{i,k-p-1:k}=(1,\textbf{d})\right\}Y_{i}}{\pi_{i,k-p-1:k}(\textbf{z})}|\mathcal{F}_{1:N,-(k-p-1:k)} \right ] \right )\\&=s(\textbf{d})\Delta^{p+1}_{HT}\left ( \mathbb{E}\left [ \frac{1}{N}\sum_{i=1}^{N}\frac{\mathbf{1}\left\{ Z_{i,k-p-1:k}=(1,\textbf{z})\right\}\mathbf{1}\left\{ D_{i,k-p-1:k}=(1,\textbf{d})\right\}Y_{i}}{\pi_{i,k-p-1:k}(1,\textbf{z})}|\mathcal{F}_{1:N,-(k-p-1:k)} \right ] \right )\\&-s(\textbf{d})\Delta^{p+1}_{HT}\left ( \mathbb{E}\left [ \frac{1}{N}\sum_{i=1}^{N}\frac{\mathbf{1}\left\{ Z_{i,k-p-1:k}=(0,\textbf{z})\right\}\mathbf{1}\left\{ D_{i,k-p-1:k}=(1,\textbf{d})\right\}Y_{i}}{\pi_{i,k-p-1:k}(0,\textbf{z})}|\mathcal{F}_{1:N,-(k-p-1:k)} \right ] \right )\\&=s(\textbf{d})\frac{1}{N}\sum_{i\in C_{k-p:k}}Y_{i,t}(d^{obs}_{i,-(k-p-1:k)},1,\textbf{d})D_{i,k-p-1}(1)-s(\textbf{d})\frac{1}{N}\sum_{i\in C_{t-p:t}}Y_{i}(d^{obs}_{i,-(k-p-1:k)},1,\textbf{d})D_{i,k-p-1}(0)\\&=s(\textbf{d})\frac{1}{N}\sum_{i\in C_{k-p-1:k}}Y_{i}(d^{obs}_{i,-(k-p-1:k)},1,\textbf{d})
\end{align*}

\noindent from which it follows that the ratio identifies

\begin{equation*}
    \frac{1}{\left | C_{k-p-1:k}\right |}\sum_{i\in C_{k-p-1:k}}Y_{i}(d^{obs}_{i,-(k-p-1:k)},1,\textbf{d})
\end{equation*}

which concludes the proof.

\subsection*{Proof of Theorem 2}

If potential outcomes are bounded, then the standard Lindeberg condition holds. The first result follows from the triangular array central limit theorem, as in \cite{bojinov}.

The variance $\left ( \sigma_{k}(1,0) \right )^{2}$ is simply

\begin{equation*}
    \left ( \sigma_{k}(1,0) \right )^{2}=\frac{1}{N^{2}}\sum_{i=1}^{N}\left ( \sigma_{i,k}(1,0) \right )^{2}
\end{equation*}

where $\left ( \sigma_{i,k}(1,0) \right )^{2}$ is defined in Auxiliary Lemma 1.

\subsection*{Proof of Theorem 3}

If potential outcomes are bounded, then the standard Lindeberg condition holds. The first result follows from the triangular array central limit theorem.

The variance is

\begin{equation*}
    \left ( \sigma_{k-p:k}(\textbf{d},\tilde{\textbf{d}}) \right )^{2}=\frac{1}{N^{2}}\sum_{i=1}^{N}\left ( \sigma_{i,k-p:k}(\textbf{d},\tilde{\textbf{d}}) \right )^{2}
\end{equation*}

See Auxiliary Lemma 2 for the necessary results.

\subsection*{Proof of Theorem 4} 

For the first part of the theorem, note that under assumptions 6-10, for all $t\in[T]$,

\begin{align*}
    &\mathbb{E}\left [ \frac{1}{N}\sum_{i=1}^{N}\frac{Z_{i,t}D_{i,t}}{\pi_{i,t}(1)}-\frac{1}{N}\sum_{i=1}^{N}\frac{(1-Z_{i,t})D_{i,t}}{\pi_{i,t}(0)}|\mathcal{F}_{1:N,t-1} \right ]\\&=\frac{1}{N}\sum_{i=1}^{N}D_{i,t}(1)-\frac{1}{N}\sum_{i=1}^{N}D_{i,t}(0)\\&=\frac{1}{N}\left | C_{t}\right |
\end{align*}

and

\begin{align*}
    &\mathbb{E}\left [ \frac{1}{N}\sum_{i=1}^{N}\frac{Z_{i,t}Y_{i,t}}{\pi_{i,t}(1)}-\frac{1}{N}\sum_{i=1}^{N}\frac{(1-Z_{i,t})Y_{i,t}}{\pi_{i,t}(0)}|\mathcal{F}_{1:N,t-1} \right ]\\&=\frac{1}{N}\sum_{i=1}^{N}Y_{i,t}(d^{obs}_{i,1:t-1},D_{i,t}(1))-\frac{1}{N}\sum_{i=1}^{N}Y_{i,t}(d^{obs}_{i,1:t-1},D_{i,t}(0))\\&=\frac{1}{N}\sum_{i\in C_{t}}\left\{Y_{i,t}(d^{obs}_{i,1:t-1},1)-Y_{i,t}(d^{obs}_{i,1:t-1},0)\right\}
\end{align*}

from which it the first result follows.

For the second part, the proof is conducted by induction. I begin by showing the result holds for $p=0$. Then, I show that if the result holds for a general $p$, then it must hold for $p+1$.

Consider the case where $p=0$ and $D_{i,t}=1$. Under Assumptions 6-10, the first stage identifies

\begin{align*}
    &\mathbb{E}\left [ \frac{1}{N}\sum_{i=1}^{N}\frac{Z_{i,t}D_{i,t}}{\pi_{i,t}(1)}-\frac{1}{N}\sum_{i=1}^{N}\frac{(1-Z_{i,t})D_{i,t}}{\pi_{i,t}(0)}|\mathcal{F}_{1:N,t-1} \right ]\\&=\frac{1}{N}\sum_{i=1}^{N}D_{i,t}(1)-\frac{1}{N}\sum_{i=1}^{N}D_{i,t}(0)=\frac{1}{N}\left | C_{t}\right |
\end{align*}

For the reduced form, note that

\begin{align*}
    &\mathbb{E}\left [ \frac{1}{N}\sum_{i=1}^{N}\frac{Z_{i,t}D_{i,t}Y_{i,t}}{\pi_{i,t}(1)}-\frac{1}{N}\sum_{i=1}^{N}\frac{(1-Z_{i,t})D_{i,t}Y_{i,t}}{\pi_{i,t}(0)}|\mathcal{F}_{1:N,t-1} \right ]\\&=\frac{1}{N}\sum_{i=1}^{N}Y_{i,t}(d^{obs}_{i,1:t-1},1)D_{i,t}(1)-\frac{1}{N}\sum_{i=1}^{N}Y_{i,t}(d^{obs}_{i,1:t-1},1)D_{i,t}(0)\\&=\frac{1}{N}\sum_{i=1}^{N}Y_{i,t}(d^{obs}_{i,1:t-1},1)(D_{i,t}(1)-D_{i,t}(0))\\&=\frac{1}{N}\sum_{i\in C_{t}} Y_{i,t}(d^{obs}_{i,1:t-1},1)
\end{align*}

and the ratio identifies

\begin{equation*}
    \frac{1}{\left | C_{t}\right |}\sum_{i\in C_{t}}Y_{i,t}(d^{obs}_{i,1:t-1},1)=m_{t}(1)
\end{equation*}

The result for the case where $D_{i,t}=0$ can be demonstrated analogously, once we note that the first stage identifies

\begin{align*}
    &\mathbb{E}\left [ \frac{1}{N}\sum_{i=1}^{N}\frac{Z_{i,t}(1-D_{i,t})}{\pi_{i,t}(1)}-\frac{1}{N}\sum_{i=1}^{N}\frac{(1-Z_{i,t})(1-D_{i,t})}{\pi_{i,t}(0)}|\mathcal{F}_{1:N,t-1} \right ]\\&=\frac{1}{N}\sum_{i=1}^{N}(1-D_{i,t}(1))-\frac{1}{N}\sum_{i=1}^{N}(1-D_{i,t}(0))\\&=-\frac{1}{N}\left | C_{t}\right |
\end{align*}

and the reduced form identifies 

\begin{align*}
    &\mathbb{E}\left [ \frac{1}{N}\sum_{i=1}^{N}\frac{Z_{i,t}(1-D_{i,t})Y_{i,t}}{\pi_{i,t}(1)}-\frac{1}{N}\sum_{i=1}^{N}\frac{(1-Z_{i,t})(1-D_{i,t})Y_{i,t}}{\pi_{i,t}(0)}|\mathcal{F}_{1:N,t-1} \right ]\\&=-\frac{1}{N}\sum_{i\in C_{t}}Y_{i,t}(d^{obs}_{i,1:t-1},0)
\end{align*}

For the general case, I prove the result only for $D_{i,t-p-1}=1$. The case for $D_{i,t-p-1}=0$ is analogous and can be derived using the intuition from the result displayed above. Let $s(\textbf{d})\in \left\{-1,1\right\}$ be a sign factor determined by the pattern of zeros and ones in $\mathbf{d}$. I begin with the first stage:

\begin{align*}
    &\Delta^{p+2}_{HT}\left ( \mathbb{E}\left [ \frac{1}{N}\sum_{i=1}^{N}\frac{\mathbf{1}\left\{ Z_{i,t-p-1:t}=\textbf{z}\right\}\mathbf{1}\left\{ D_{i,t-p-1:t}=(1,\textbf{d})\right\}}{\pi_{i,t-p-1:t}(\textbf{z})}|\mathcal{F}_{1:N,t-p-2} \right ] \right )\\&=s(\textbf{d})\Delta^{p+1}_{HT}\left ( \mathbb{E}\left [ \frac{1}{N}\sum_{i=1}^{N}\frac{\mathbf{1}\left\{ Z_{i,t-p-1:t}=(1,\textbf{z})\right\}\mathbf{1}\left\{ D_{i,t-p-1:t}=(1,\textbf{d})\right\}}{\pi_{i,t-p-1:t}(1,\textbf{z})}|\mathcal{F}_{1:N,t-p-2} \right ] \right )\\&-s(\textbf{d})\Delta^{p+1}_{HT}\left ( \mathbb{E}\left [ \frac{1}{N}\sum_{i=1}^{N}\frac{\mathbf{1}\left\{ Z_{i,t-p-1:t}=(0,\textbf{z})\right\}\mathbf{1}\left\{ D_{i,t-p-1:t}=(1,\textbf{d})\right\}}{\pi_{i,t-p-1:t}(0,\textbf{z})}|\mathcal{F}_{1:N,t-p-2} \right ] \right )\\&=s(\textbf{d})\frac{1}{N}\sum_{i=1}^{N}D_{i,t-p-1}(1)\mathbf{1}\left\{ i\in C_{t-p:t}\right\}-s(\textbf{d})\frac{1}{N}\sum_{i=1}^{N}D_{i,t-p-1}(0)\mathbf{1}\left\{ i\in C_{t-p:t}\right\}\\&=s(\textbf{d})\frac{1}{N}\sum_{i=1}^{N}\mathbf{1}\left\{ i\in C_{t-p-1:t}\right\}=s(\textbf{d})\frac{1}{N}\left | C_{t-p-1:t}\right |
\end{align*}

\noindent where the first equality follows from the definition of $\Delta^{p+2}(.)$, the second equality follows from Assumptions 9 and 10, and assuming the result holds for $p$ and the third equality follows from Assumptions 7 (ii) and Assumption 8.

For the modified reduced form, note that 

\begin{align*}
    &\Delta^{p+2}_{HT}\left ( \mathbb{E}\left [ \frac{1}{N}\sum_{i=1}^{N}\frac{\mathbf{1}\left\{ Z_{i,t-p-1:t}=\textbf{z}\right\}\mathbf{1}\left\{ D_{i,t-p-1:t}=(1,\textbf{d})\right\}Y_{i,t}}{\pi_{i,t-p-1:t}(\textbf{z})}|\mathcal{F}_{1:N,t-p-2} \right ] \right )\\&=s(\textbf{d})\Delta^{p+1}_{HT}\left ( \mathbb{E}\left [ \frac{1}{N}\sum_{i=1}^{N}\frac{\mathbf{1}\left\{ Z_{i,t-p-1:t}=(1,\textbf{z})\right\}\mathbf{1}\left\{ D_{i,t-p-1:t}=(1,\textbf{d})\right\}Y_{i,t}}{\pi_{i,t-p-1:t}(1,\textbf{z})}|\mathcal{F}_{1:N,t-p-2} \right ] \right )\\&-s(\textbf{d})\Delta^{p+1}_{HT}\left ( \mathbb{E}\left [ \frac{1}{N}\sum_{i=1}^{N}\frac{\mathbf{1}\left\{ Z_{i,t-p-1:t}=(0,\textbf{z})\right\}\mathbf{1}\left\{ D_{i,t-p-1:t}=(1,\textbf{d})\right\}Y_{i,t}}{\pi_{i,t-p-1:t}(0,\textbf{z})}|\mathcal{F}_{1:N,t-p-2} \right ] \right )\\&=s(\textbf{d})\frac{1}{N}\sum_{i\in C_{t-p:t}}Y_{i,t}(d^{obs}_{i,1:t-p-2},1,\textbf{d})D_{i,t-p-1}(1)-s(\textbf{d})\frac{1}{N}\sum_{i\in C_{t-p:t}}Y_{i,t}(d^{obs}_{i,1:t-p-2},1,\textbf{d})D_{i,t-p-1}(0)\\&=s(\textbf{d})\frac{1}{N}\sum_{i\in C_{t-p-1:t}}Y_{i,t}(d^{obs}_{i,1:t-p-2},1,\textbf{d})
\end{align*}

\noindent from which it follows that the ratio identifies

\begin{equation*}
    \frac{1}{\left | C_{t-p-1:t}\right |}\sum_{i\in C_{t-p-1:t}}Y_{i,t}(d^{obs}_{i,1:t-p-2},1,\textbf{d})
\end{equation*}

which concludes the proof.

\subsection*{Proof of Theorem 5}

If potential outcomes are bounded, then the standard Lindeberg condition holds. The first result follows from the triangular array central limit theorem, as in \cite{bojinov}.

The variance $\left ( \sigma_{t}(1,0;0) \right )^{2}$ is simply

\begin{equation*}
    \left ( \sigma_{t}(1,0;0) \right )^{2}=\frac{1}{N^{2}}\sum_{i=1}^{N}\left ( \sigma_{i,t}(1,0;0) \right )^{2}
\end{equation*}

where $\left ( \sigma_{i,t}(1,0;0) \right )^{2}$ is defined in Lemma 3 from Appendix B.

\subsection*{Proof of Theorem 6}

The result follows the same reasoning as the one in Theorem 5. This time, however, we have

\begin{equation*}
    \left ( \sigma_{t}(\textbf{d},\tilde{\textbf{d}};p) \right )^{2}=\frac{1}{N^{2}}\sum_{i=1}^{N}\left ( \sigma_{i,t}(\textbf{d},\tilde{\textbf{d}};p) \right )^{2}
\end{equation*}

See Lemma 4 from Appendix B for the necessary results.

\section*{Appendix B - Auxiliary Lemmas}

\begin{lemma}
    Suppose that potential outcomes are bounded and that Assumptions 1-5 hold. Then,

    \begin{equation*}
    \mathbb{V}\left [ u_{i,k}|\mathcal{F}_{i,-k} \right ]\approx \frac{\left ( \sigma_{i,k}^{RF}(1,0) \right )^{2}}{\left ( \tau_{i,k}^{FS}(1,0) \right )^{2}}+\frac{\left ( \tau_{i,k}^{RF}(1,0) \right )^2\left ( \sigma_{i,k}^{FS}(1,0) \right )^{2}}{\left ( \tau_{i,k}^{FS}(1,0) \right )^{4}}-\frac{2\tau_{i,k}^{RF}(1,0)\mathbb{C}ov\left ( \widehat{\tau}_{i,k}^{RF}(1,0),\widehat{\tau}_{i,k}^{FS}(1,0) \right|\mathcal{F}_{i,-k} )}{\left ( \tau_{i,k}^{FS}(1,0) \right )^{3}}
\end{equation*}

\end{lemma}

\subsection*{Proof of Lemma 1}

Define $W_{i,k-p:k}(\textbf{z})=\pi_{i,k-p:k}(\textbf{z})^{-1}\mathbf{1}\left \{ Z_{i,k-p:k}=\textbf{z} \right \}$. We have

 \begin{align*}
     &\mathbb{E}\left [ W_{i,k-p:k}(\textbf{z})|\mathcal{F}_{i,-(k-p)} \right ]=1,\\&
     \mathbb{V}\left [ W_{i,k-p:k}(\textbf{z})|\mathcal{F}_{i,-(k-p)}  \right ]=\pi_{i,t-p:t}(\textbf{z})^{-1}(1-\pi_{i,k-p:k}(\textbf{z})),\\&
     \mathbb{C}ov\left [ W_{i,k-p:k}(\textbf{z}),W_{i,k-p:l}(\tilde{\textbf{z}})|\mathcal{F}_{i,-(k-p)}  \right ]=-1
 \end{align*}

I analyze the properties of the estimator for each stage separately before analyzing the properties of the two-stage estimator. I begin with the first stage. Define

\begin{equation*}
     u_{i,k}^{FS}=\widehat{\tau}_{i,k}^{FS}-\tau_{i,k}^{FS}=D_{i,k}(1)(W_{i,k}(1)-1)-D_{i,k}(0)(W_{i,k}(0)-1)
\end{equation*}

 We have $\mathbb{E}\left [ u_{i,k}^{FS}|\mathcal{F}_{i,-k}  \right ]=0$. Now, let's look at the variance:

 \begin{align*}
     &\mathbb{V}\left [ u_{i,k}^{FS}|\mathcal{F}_{i,-k}  \right ]=D_{i,k}(1)^{2}\mathbb{V}\left [ W_{i,k}(1) \right ]+D_{i,k}(0)^{2}\mathbb{V}\left [ W_{i,k}(0) \right ]\\&-2D_{i,k}(1)D_{i,k}(0)\mathbb{C}ov\left [W_{i,k}(1),W_{i,k}(0)  \right ]\\&=D_{i,k}(1)^{2}\pi_{i,k}(1)^{-1}(1-\pi_{i,k}(1))+D_{i,k}(0)^{2}\pi_{i,k}(0)^{-1}(1-\pi_{i,k}(0))\\&+2D_{i,k}(1)D_{i,k}(0)\\&=\frac{D_{i,k}(1)^{2}}{\pi_{i,k}(1)}+\frac{D_{i,k}(0)^{2}}{\pi_{i,t}(0)}-(D_{i,k}(1)-D_{i,k}(0))^{2}\\&=\left ( \gamma_{i,k}^{FS}(1,0) \right )^{2}-(D_{i,k}(1)-D_{i,k}(0))^{2}=\left ( \sigma_{i,k}^{FS}(1,0) \right )^{2}
 \end{align*}

Now, consider the reduce form. Define

\begin{equation*}
    u_{i,k}^{RF}=\widehat{\tau}_{i,k}^{RF}-\tau_{i,k}^{RF}=Y_{i}(d^{obs}_{i,-k},D_{i,k}(1))(W_{i,k}(1)-1)-Y_{i}(d^{obs}_{i,-k},D_{i,k}(0))(W_{i,k}(0)-1)
\end{equation*}

We have $\mathbb{E}\left [ u_{i,k}^{RF}|\mathcal{F}_{i,-k}  \right ]=0$. Now, let's look at the variance:

\begin{align*}
    &\mathbb{V}\left [ u_{i,k}^{RF}|\mathcal{F}_{i,-k}  \right ]=Y_{i}(d^{obs}_{i,-k},D_{i,k}(1))^{2}\mathbb{V}\left [ W_{i,k}(1) \right ]+Y_{i}(d^{obs}_{i,-k},D_{i,k}(0))^{2}\mathbb{V}\left [ W_{i,k}(0) \right ]\\&-2Y_{i}(d^{obs}_{i,-k},D_{i,k}(1))Y_{i}(d^{obs}_{i,-k},D_{i,k}(0))\mathbb{C}ov\left [ W_{i,k}(1),W_{i,k}(0) \right ]\\&=Y_{i}(d^{obs}_{i,-k},D_{i,k}(1))^{2}\pi_{i,k}(1)^{-1}(1-\pi_{i,k}(1))+Y_{i}(d^{obs}_{i,-k},D_{i,k}(0))^{2}\pi_{i,k}(0)^{-1}(1-\pi_{i,k}(0))\\&+2Y_{i}(d^{obs}_{i,-k},D_{i,k}(1))Y_{i}(d^{obs}_{i,-k},D_{i,k}(0))\\&=\frac{Y_{i}(d^{obs}_{i,-k},D_{i,k}(1))^{2}}{\pi_{i,k}(1)}+\frac{Y_{i}(d^{obs}_{i,-k},D_{i,k}(0))^{2}}{\pi_{i,k}(0)}-(Y_{i}(d^{obs}_{i,-k},D_{i,k}(1))-Y_{i}(d^{obs}_{i,-k},D_{i,k}(0)))^{2}\\&=\left ( \gamma_{i,k}^{RF}(1,0) \right )^{2}-(Y_{i}(d^{obs}_{i,-k},D_{i,k}(1))-Y_{i}(d^{obs}_{i,-k},D_{i,k}(0)))^{2}=\left ( \sigma_{i,k}^{RF}(1,0) \right )^{2}
\end{align*}

From the results above, it follows that

\begin{equation*}
    \frac{\mathbb{E}\left [ u_{i,k}^{RF}|\mathcal{F}_{i,-k}  \right ]}{\mathbb{E}\left [ u_{i,k}^{FS}|\mathcal{F}_{i,-k}  \right ]}=0
\end{equation*}

We apply the Uniform Delta Method to obtain

\begin{align*}
    &\mathbb{V}\left [ u_{i,k}|\mathcal{F}_{i,-k} \right ]=g^{'}\begin{pmatrix}
\left ( \sigma_{i,k}^{RF}(1,0) \right )^{2}\\ 
\left ( \sigma_{i,k}^{FS}(1,0) \right )^{2}
\end{pmatrix}g:=\left ( \sigma_{i,k}(1,0) \right )^{2}\\&\approx \frac{\left ( \sigma_{i,k}^{RF}(1,0) \right )^{2}}{\left ( \tau_{i,k}^{FS}(1,0) \right )^{2}}+\frac{\left ( \tau_{i,k}^{RF}(1,0) \right )^2\left ( \sigma_{i,k}^{FS}(1,0) \right )^{2}}{\left ( \tau_{i,k}^{FS}(1,0) \right )^{4}}-\frac{2\tau_{i,k}^{RF}(1,0)\mathbb{C}ov\left ( \widehat{\tau}_{i,k}^{RF}(1,0),\widehat{\tau}_{i,k}^{FS}(1,0) \right|\mathcal{F}_{i,-k} )}{\left ( \tau_{i,k}^{FS}(1,0) \right )^{3}}
\end{align*}

Where $g$ is the gradient of $h(x,y)=x/y$ evaluated at $(\tau_{i,k}^{RF},\tau_{i,k}^{FS})$, which concludes the proof.

\begin{lemma}
    Suppose that potential outcomes are bounded and that Assumptions 1-5 hold. Then, 

    {\scriptsize \begin{equation*}
    \mathbb{V}\left [ u_{i,k-p:k}(\textbf{d})|\mathcal{F}_{i,-(k-p:k)}  \right ]\approx \frac{\left ( \sigma_{i,k-p:k}^{RF}(\textbf{d}) \right )^{2}}{\left ( m_{i,k-p:k}^{FS}(\textbf{d}) \right )^{2}}+\frac{\left ( m_{i,k-p:k}^{RF}(\textbf{d}) \right )^2\left ( \sigma_{i,k-p:k}^{FS}(\textbf{d}) \right )^{2}}{\left ( m_{i,k-p:k}^{FS}(\textbf{d}) \right )^{4}}-\frac{2m_{i,k-p:k}^{RF}(\textbf{d})\mathbb{C}ov\left ( \widehat{m}_{i,k-p:k}^{RF}(\textbf{d}),\widehat{m}_{i,k-p:k}^{FS}(\textbf{d}) \right )}{\left ( m_{i,k-p:k}^{FS}(\textbf{d}) \right )^{3}}
\end{equation*}}
\end{lemma}

\subsection*{Proof of Lemma 2}

Define

\begin{align*}
     &u_{i,k-p:k}^{FS}(\textbf{d})=\widehat{m}_{i,k-p:k}^{FS}(\textbf{d})-m_{i,k-p:k}^{FS}(\textbf{d})=\Delta^{p+1}\left ( \mathbf{1}\left \{ D_{i,k-p:k}(\textbf{z}) \right \}(W_{i,k-p:k}(\textbf{z})-1) \right ),\\&u_{i,k-p:k}^{RF}(\textbf{d})=\widehat{m}_{i,k-p:k}^{RF}(\textbf{d})-m_{i,k-p:k}^{RF}(\textbf{d})=\Delta^{p+1}\left ( Y_{i}(d^{obs}_{i,-(k-p:k},\textbf{d})\mathbf{1}\left \{ D_{i,k-p:k}(\textbf{z}) \right \}(W_{i,k-p:k}(\textbf{z})-1) \right )
 \end{align*}

We have $\mathbb{E}\left [ u_{i,k-p:k}^{FS}(\textbf{d})|\mathcal{F}_{i,-(k-p:k)}  \right ]=\mathbb{E}\left [ u_{i,k-p:k}^{RF}(\textbf{d})|\mathcal{F}_{i,-(k-p:k)}  \right ]=0$ for all $\textbf{d}\in\left \{ 0,1 \right \}^{p+1}$. Furthermore, define $u_{i,k-p:k}^{RF}(\textbf{d},\tilde{\textbf{d}})=u_{i,k-p:k}^{RF}(\textbf{d})+u^{RF}_{i,k-p:k}(\tilde{\textbf{d}})$ and $u_{i,k-p:k}^{FS}(\textbf{d},\tilde{\textbf{d}})$ analogously. It follows that $\mathbb{E}\left [ u^{RF}_{i,k-p:k}(\textbf{d},\tilde{\textbf{d}};p)|\mathcal{F}_{i,-(k-p:k)}  \right ]=\mathbb{E}\left [ u^{FS}_{i,k-p:k}(\textbf{d},\tilde{\textbf{d}})|\mathcal{F}_{i,-(k-p:k)}  \right ]=0$.

Now let's consider the variance. I prove the result by induction for the reduced form and first stage of each potential outcome separately, and build on it to derive the properties of the estimator.

First, let's consider the case when $p=0$ and $D_{i,k}=1$ (the case for $D_{i,k}=0$ is analogous). For the first stage, 

 \begin{align*}
     &\mathbb{V}\left [ u_{i,k}^{FS}(1)|\mathcal{F}_{i,-k}  \right ]=D_{i,k}(1)^{2}\mathbb{V}\left [ W_{i,k}(1) \right ]+D_{i,k}(0)^{2}\mathbb{V}\left [ W_{i,k}(0) \right ]\\&-2D_{i,k}(1)D_{i,k}(0)\mathbb{C}ov\left [W_{i,k}(1),W_{i,k}(0)  \right ]\\&=D_{i,k}(1)^{2}\pi_{i,k}(1)^{-1}(1-\pi_{i,k}(1))+D_{i,k}(0)^{2}\pi_{i,k}(0)^{-1}(1-\pi_{i,k}(0))\\&+2D_{i,k}(1)D_{i,k}(0)\\&=\frac{D_{i,k}(1)^{2}}{\pi_{i,k}(1)}+\frac{D_{i,k}(0)^{2}}{\pi_{i,k}(0)}-(D_{i,k}(1)-D_{i,k}(0))^{2}\\&=\left ( \gamma_{i,k}^{FS}(1) \right )^{2}-(D_{i,k}(1)-D_{i,k}(0))^{2}=\left ( \sigma_{i,k}^{FS}(1) \right )^{2}
 \end{align*}

Now, consider the reduced form:

\begin{align*}
    &\mathbb{V}\left [ u_{i,k}^{RF}(1)|\mathcal{F}_{i,-k}  \right ]=\left ( Y_{i}(d^{obs}_{i,-k},1)D_{i,k}(1) \right )^{2}\mathbb{V}\left [ W_{i,k}(1) \right ]+\left ( Y_{i}(d^{obs}_{i,-k},1)D_{i,k}(0) \right )^{2}\mathbb{V}\left [ W_{i,k}(0) \right ]\\&-2Y_{i}(d^{obs}_{i,-k},1)D_{i,k}(1)Y_{i}(d^{obs}_{i,-k},1)D_{i,k}(0)\mathbb{C}ov\left [ W_{i,k}(1),W_{i,k}(0) \right ]\\&=\frac{\left ( Y_{i}(d^{obs}_{i,-k},1)D_{i,k}(1) \right )^{2}}{\pi_{i,k}(1)}+\frac{\left ( Y_{i}(d^{obs}_{i,-k},1)D_{i,k}(0) \right )^{2}}{\pi_{i,k}(0)}\\&-(Y_{i}(d^{obs}_{i,-k},1)D_{i,k}(1)-Y_{i}(d^{obs}_{i,-k},1)D_{i,k}(0))^{2}\\&=\left ( \gamma_{i,k}^{RF}(1) \right )^{2}-(Y_{i}(d^{obs}_{i,-k},1)D_{i,k}(1)-Y_{i}(d^{obs}_{i,-k},1)D_{i,k}(0))^{2}=\left ( \sigma_{i,k}^{RF}(1) \right )^{2}
\end{align*}

and the result follows from the Delta method.

Now, I proceed by deriving the expression for a general $p+1$, assuming that the result holds for $p$. For the first stage, 

\begin{align*}
    &\mathbb{V}\left [ u_{i,k-p-1:k}^{FS}(\textbf{d})|\mathcal{F}_{i,-(k-p-1)}  \right ]=\mathbb{V}\left [ \Delta^{p+2}\left (\mathbf{1}\left \{ D_{i,k-p-1:k}(\textbf{z})=\textbf{d} \right \}(W_{i,k-p-1:k}(\textbf{z})-1)  \right ) \right ]\\&=\mathbb{V} [ \Delta^{p+1}\left (\mathbf{1}\left \{ D_{i,k-p-1:k}(1,\textbf{z})=\textbf{d} \right \}(W_{i,k-p-1:k}(1,\textbf{z})-1)  \right ) \\&-\Delta^{p+1}\left (\mathbf{1}\left \{ D_{i,k-p-1:k}(0,\textbf{z})=\textbf{d} \right \}(W_{i,k-p-1:k}(0,\textbf{z})-1)  \right ) ]\\&=\mathbb{V}\left [ \Delta^{p+1}\left (\mathbf{1}\left \{ D_{i,k-p-1:k}(1,\textbf{z})=\textbf{d} \right \}(W_{i,k-p-1:k}(1,\textbf{z})-1)  \right )  \right ]\\&+\mathbb{V}\left [ \Delta^{p+1}\left (\mathbf{1}\left \{ D_{i,k-p-1:k}(0,\textbf{z})=\textbf{d} \right \}(W_{i,k-p-1:k}(0,\textbf{z})-1)  \right )  \right ]\\&-2 \Delta^{p+1}\left (\mathbf{1}\left \{ D_{i,k-p-1:k}(1,\textbf{z})=\textbf{d} \right \}\right)\Delta^{p+1}\left (\mathbf{1}\left \{ D_{i,k-p-1:k}(0,\textbf{z})=\textbf{d} \right \}\right)\mathbb{C}ov\left [ W_{i,k-p-1:k}(1,\textbf{z}),W_{i,k-p-1:k}(0,\textbf{z}) \right ]\\&=\sum_{\textbf{z}\in\left \{ 0,1 \right \}^{p+1}}\frac{\mathbf{1}\left \{ D_{i,k-p-1:k}(1,\textbf{z}) \right \}=\textbf{d}}{\pi_{i,k-p-1:k}(1,\textbf{z})}-\Delta^{p+1}\left (  D_{i,k-p-1:k}(1,\textbf{z})\right )^{2}\\&+\sum_{\textbf{z}\in\left \{ 0,1 \right \}^{p+1}}\frac{\mathbf{1}\left \{ D_{i,k-p-1:k}(0,\textbf{z}) \right \}=\textbf{d}}{\pi_{i,k-p-1:k}(0,\textbf{z})}-\Delta^{p+1}\left (  D_{i,k-p-1:k}(0,\textbf{z})\right )^{2}\\&+2\Delta^{p+1}\left (D_{i,k-p-1:k}(1,\textbf{z})  \right )\Delta^{p+1}\left ( D_{i,k-p-1:k}(0,\textbf{z}) \right )\\&=\sum_{\textbf{z}\in\left \{ 0,1 \right \}^{p+2}}\frac{\mathbf{1}\left \{ D_{i,k-p-1:k}(\textbf{z}) \right \}=\textbf{d}}{\pi_{i,k-p-1:k}(\textbf{z})}\\&-\left ( \Delta^{p+2}\left ( \mathbf{1}\left \{ D_{i,k-p-1:k}(1,\textbf{z}) \right \}=\textbf{d} \right )-\Delta^{p+2}\left ( \mathbf{1}\left \{ D_{i,k-p-1:k}(0,\textbf{z}) \right \}=\textbf{d} \right ) \right )^{2}\\&=\left ( \gamma_{i,k-p-1:k}^{FS}(\textbf{d}) \right )^{2}-\left ( \Delta^{p+2}\left ( \mathbf{1}\left \{ D_{i,k-p-1:k}(\textbf{z}) \right \}=\textbf{d} \right ) \right )^{2}\\&=\left ( \sigma_{i,k-p-1:k}^{FS}(\textbf{d}) \right )^{2}
\end{align*}

Analogously, for the reduced form it can be shown that

\begin{align*}
    &\mathbb{V}\left [ u_{ik-p-1:k}^{RF}(\textbf{d})|\mathcal{F}_{i,-(k-p-1)}  \right ]=\sum_{\textbf{z}\in\left \{ 0,1 \right \}^{p+2}}\frac{\left ( Y_{i}(d^{obs}_{i,-(k-p-1:k)},\textbf{d})\mathbf{1}\left \{ D_{i,k-p-1:k}(\textbf{z}) =\textbf{d}\right \} \right )^{2}}{\pi_{i,k-p-1:k}(\textbf{z})}\\&-\left ( \Delta^{p+2}\left ( Y_{i,t}(d^{obs}_{i,-(k-p-1:k)},\textbf{d})\mathbf{1}\left \{ D_{i,k-p-1:k}(\textbf{z}) =\textbf{d}\right \} \right ) \right )^{2}\\&=\left ( \gamma_{i,k-p-1:k}^{RF}(\textbf{d}) \right )^{2}-\left ( \Delta^{p+2}\left ( Y_{i}(d^{obs}_{i,-(k-p-1:k)},\textbf{d})\mathbf{1}\left \{ D_{i,k-p-1:k}(\textbf{z}) =\textbf{d}\right \} \right ) \right )^{2}=\left ( \sigma_{i,k-p-1:k}^{RF}(\textbf{d}) \right )^{2}
\end{align*}

From the results above, we have 
\begin{align*}
    &\mathbb{V}\left [ u_{i,k-p:k}(\textbf{d})|\mathcal{F}_{i,-(k-p:k)}  \right ]=g^{'}\begin{pmatrix}
\left ( \sigma_{i,k-p:k}^{RF}(\textbf{d}) \right )^{2}\\ 
\left ( \sigma_{i,k-p:k}^{FS}(\textbf{d}) \right )^{2}
\end{pmatrix}g=\left ( \sigma_{i,k-p:k}(\textbf{d}) \right )^{2}\\&\approx \frac{\left ( \sigma_{i,k-p:k}^{RF}(\textbf{d}) \right )^{2}}{\left ( m_{i,k-p:k}^{FS}(\textbf{d}) \right )^{2}}+\frac{\left ( m_{i,k-p:k}^{RF}(\textbf{d}) \right )^2\left ( \sigma_{i,k-p:k}^{FS}(\textbf{d}) \right )^{2}}{\left ( m_{i,k-p:k}^{FS}(\textbf{d}) \right )^{4}}-\frac{2m_{i,k-p:k}^{RF}(\textbf{d})\mathbb{C}ov\left ( \widehat{m}_{i,k-p:k}^{RF}(\textbf{d}),\widehat{m}_{i,k-p:k}^{FS}(\textbf{d}) \right )}{\left ( m_{i,k-p:k}^{FS}(\textbf{d}) \right )^{3}}
\end{align*}

Where $g$ is the gradient of $h(x,y)=x/y$ evaluated at $(m_{i,k-p:k}^{RF}(\textbf{d}),m_{i,k-p:k}^{FS}(\textbf{d}))$ 

The variance for $\widehat{\tau}_{i,k-p:k}(\textbf{d},\tilde{\textbf{d}})$ is simply $\left ( \sigma_{i,k-p:k}(\textbf{d}) \right )^{2}+\left ( \sigma_{i,k-p:k}(\tilde{\textbf{d}}) \right )^{2}$, which concludes the proof.

\begin{lemma}
    Suppose that potential outcomes are bounded and that Assumptions 6-10 hold. Then, 

{\scriptsize\begin{equation*}
    \mathbb{V}\left [ u_{i,t}|\mathcal{F}_{i,t-1} \right ]=\approx \frac{\left ( \sigma_{i,t}^{RF}(1,0;0) \right )^{2}}{\left ( \tau_{i,t}^{FS}(1,0;0) \right )^{2}}+\frac{\left ( \tau_{i,t}^{RF}(1,0;0) \right )^2\left ( \sigma_{i,t}^{FS}(1,0;0) \right )^{2}}{\left ( \tau_{i,t}^{FS}(1,0;0) \right )^{4}}-\frac{2\tau_{i,t}^{RF}(1,0;0)\mathbb{C}ov\left ( \widehat{\tau}_{i,t}^{RF}(1,0;0),\widehat{\tau}_{i,t}^{FS}(1,0;0) \right|\mathcal{F}_{i,t-1} )}{\left ( \tau_{i,t}^{FS}(1,0;0) \right )^{3}}
\end{equation*}}

\end{lemma}

\subsection*{Proof of Lemma 3}

Define $W_{i,t-p:t}(\textbf{z})=\pi_{i,t-p:t}(\textbf{z})^{-1}\mathbf{1}\left \{ Z_{i,t-p:t}=\textbf{z} \right \}$. From Lemma A.1 in \cite{bojinov}, we have

 \begin{align*}
     &\mathbb{E}\left [ W_{i,t-p:t}(\textbf{z})|\mathcal{F}_{i,t-p-1} \right ]=1,\\&
     \mathbb{V}\left [ W_{i,t-p:t}(\textbf{z})|\mathcal{F}_{i,t-p-1}  \right ]=\pi_{i,t-p:t}(\textbf{z})^{-1}(1-\pi_{i,t-p:t}(\textbf{z})),\\&
     \mathbb{C}ov\left [ W_{i,t-p:t}(\textbf{z}),W_{i,t-p:t}(\tilde{\textbf{z}})|\mathcal{F}_{i,t-p-1}  \right ]=-1
 \end{align*}

 I analyze the properties of the estimator for each stage separately before analyzing the properties of the two-stage estimator. I begin with the first stage. Define

\begin{equation*}
     u_{i,t}^{FS}=\widehat{\tau}_{i,t}^{FS}-\tau_{i,t}^{FS}=D_{i,t}(1)(W_{i,t}(1)-1)-D_{i,t}(0)(W_{i,t}(0)-1)
\end{equation*}

 Lemma A.1 from \cite{bojinov} implies that $\mathbb{E}\left [ u_{i,t}^{FS}|\mathcal{F}_{i,t-p-1}  \right ]=0$. Hence, the error terms from the estimator are a martingale difference sequence and so, uncorrelated through time. Now, let's look at the variance:

 \begin{align*}
     &\mathbb{V}\left [ u_{i,t}^{FS}|\mathcal{F}_{i,t-1}  \right ]=D_{i,t}(1)^{2}\mathbb{V}\left [ W_{i,t}(1) \right ]+D_{i,t}(0)^{2}\mathbb{V}\left [ W_{i,t}(0) \right ]\\&-2D_{i,t}(1)D_{i,t}(0)\mathbb{C}ov\left [W_{i,t}(1),W_{i,t}(0)  \right ]\\&=D_{i,t}(1)^{2}\pi_{i,t}(1)^{-1}(1-\pi_{i,t}(1))+D_{i,t}(0)^{2}\pi_{i,t}(0)^{-1}(1-\pi_{i,t}(0))\\&+2D_{i,t}(1)D_{i,t}(0)\\&=\frac{D_{i,t}(1)^{2}}{\pi_{i,t}(1)}+\frac{D_{i,t}(0)^{2}}{\pi_{i,t}(0)}-(D_{i,t}(1)-D_{i,t}(0))^{2}\\&=\left ( \gamma_{i,t}^{FS}(1,0)(0) \right )^{2}-(D_{i,t}(1)-D_{i,t}(0))^{2}=\left ( \sigma_{i,t}^{FS}(1,0)(0) \right )^{2}
 \end{align*}

Now, consider the reduce form. Define

\begin{equation*}
    u_{i,t}^{RF}=\widehat{\tau}_{i,t}^{RF}-\tau_{i,t}^{RF}=Y_{i,t}(d^{obs}_{i,1:t-1},D_{i,t}(1))(W_{i,t}(1)-1)-Y_{i,t}(d^{obs}_{i,1:t-1},D_{i,t}(0))(W_{i,t}(0)-1)
\end{equation*}

We have $\mathbb{E}\left [ u_{i,t}^{RF}|\mathcal{F}_{i,t-1}  \right ]=0$. Hence, the error terms from the estimator are a martingale difference sequence and so uncorrelated through time. Now, let's look at the variance:

\begin{align*}
    &\mathbb{V}\left [ u_{i,t}^{RF}|\mathcal{F}_{i,t-1}  \right ]=Y_{i,t}(d^{obs}_{i,1:t-1},D_{i,t}(1))^{2}\mathbb{V}\left [ W_{i,t}(1) \right ]+Y_{i,t}(d^{obs}_{i,1:t-1},D_{i,t}(0))^{2}\mathbb{V}\left [ W_{i,t}(0) \right ]\\&-2Y_{i,t}(d^{obs}_{i,1:t-1},D_{i,t}(1))Y_{i,t}(d^{obs}_{i,1:t-1},D_{i,t}(0))\mathbb{C}ov\left [ W_{i,t}(1),W_{i,t}(0) \right ]\\&=Y_{i,t}(d^{obs}_{i,1:t-1},D_{i,t}(1))^{2}\pi_{i,t}(1)^{-1}(1-\pi_{i,t}(1))+Y_{i,t}(d^{obs}_{i,1:t-1},D_{i,t}(0))^{2}\pi_{i,t}(0)^{-1}(1-\pi_{i,t}(0))\\&+2Y_{i,t}(d^{obs}_{i,1:t-1},D_{i,t}(1))Y_{i,t}(d^{obs}_{i,1:t-1},D_{i,t}(0))\\&=\frac{Y_{i,t}(d^{obs}_{i,1:t-1},D_{i,t}(1))^{2}}{\pi_{i,t}(1)}+\frac{Y_{i,t}(d^{obs}_{i,1:t-1},D_{i,t}(0))^{2}}{\pi_{i,t}(0)}-(Y_{i,t}(d^{obs}_{i,1:t-1},D_{i,t}(1))-Y_{i,t}(d^{obs}_{i,1:t-1},D_{i,t}(0)))^{2}\\&=\left ( \gamma_{i,t}^{RF}(1,0;0) \right )^{2}-(Y_{i,t}(d^{obs}_{i,1:t-1},D_{i,t}(1))-Y_{i,t}(d^{obs}_{i,1:t-1},D_{i,t}(0)))^{2}=\left ( \sigma_{i,t}^{RF}(1,0;0) \right )^{2}
\end{align*}

From the results above, it follows that

\begin{equation*}
    \frac{\mathbb{E}\left [ u_{i,t}^{RF}|\mathcal{F}_{i,t-1}  \right ]}{\mathbb{E}\left [ u_{i,t}^{FS}|\mathcal{F}_{i,t-1}  \right ]}=0
\end{equation*}

We apply the Uniform Delta Method to obtain

\begin{align*}
    &\mathbb{V}\left [ u_{i,t}|\mathcal{F}_{i,t-1} \right ]=g^{'}\begin{pmatrix}
\left ( \sigma_{i,t}^{RF}(1,0;0) \right )^{2}\\ 
\left ( \sigma_{i,t}^{FS}(1,0;0) \right )^{2}
\end{pmatrix}g:=\left ( \sigma_{i,t}(1,0;0) \right )^{2}\\&\approx \frac{\left ( \sigma_{i,t}^{RF}(1,0;0) \right )^{2}}{\left ( \tau_{i,t}^{FS}(1,0;0) \right )^{2}}+\frac{\left ( \tau_{i,t}^{RF}(1,0;0) \right )^2\left ( \sigma_{i,t}^{FS}(1,0;0) \right )^{2}}{\left ( \tau_{i,t}^{FS}(1,0;0) \right )^{4}}-\frac{2\tau_{i,t}^{RF}(1,0;0)\mathbb{C}ov\left ( \widehat{\tau}_{i,t}^{RF}(1,0;0),\widehat{\tau}_{i,t}^{FS}(1,0;0) \right|\mathcal{F}_{i,t-1} )}{\left ( \tau_{i,t}^{FS}(1,0;0) \right )^{3}}
\end{align*}

Where $g$ is the gradient of $h(x,y)=x/y$ evaluated at $(\tau_{i,t}^{RF},\tau_{i,t}^{FS})$, which concludes the proof.

\begin{lemma}
    Suppose that potential outcomes are bounded and Assumptions 6-10 hold. Then,

    {\scriptsize\begin{equation*}
    \mathbb{V}\left [ u_{i,t}(\textbf{d})|\mathcal{F}_{i,t-p-1}  \right ]\approx \frac{\left ( \sigma_{i,t}^{RF}(\textbf{d}) \right )^{2}}{\left ( m_{i,t}^{FS}(\textbf{d}) \right )^{2}}+\frac{\left ( m_{i,t}^{RF}(\textbf{d}) \right )^2\left ( \sigma_{i,t}^{FS}(\textbf{d}) \right )^{2}}{\left ( m_{i,t}^{FS}(\textbf{d}) \right )^{4}}-\frac{2m_{i,t}^{RF}(\textbf{d})\mathbb{C}ov\left ( \widehat{m}_{i,t}^{RF}(\textbf{d}),\widehat{m}_{i,t}^{FS}(\textbf{d}) |\mathcal{F}_{i,t-p-1}\right )}{\left ( m_{i,t}^{FS}(\textbf{d}) \right )^{3}}
\end{equation*}}

\end{lemma}

\subsection*{Proof of Lemma 4}

Define 

 \begin{align*}
     &u_{i,t}^{FS}(\textbf{d})=\widehat{m}_{i,t}^{FS}(\textbf{d})-m_{i,t}^{FS}(\textbf{d})=\Delta^{p+1}\left ( \mathbf{1}\left \{ D_{i,t-p:t}(\textbf{z}) \right \}(W_{i,t-p:t}(\textbf{z})-1) \right ),\\&u_{i,t}^{RF}(\textbf{d})=\widehat{m}_{i,t}^{RF}(\textbf{d})-m_{i,t}^{RF}(\textbf{d})=\Delta^{p+1}\left ( Y_{i,t}(d^{obs}_{i,1:t-p-1},\textbf{d})\mathbf{1}\left \{ D_{i,t-p:t}(\textbf{z}) \right \}(W_{i,t-p:t}(\textbf{z})-1) \right )
 \end{align*}

We have $\mathbb{E}\left [ u_{i,t}^{FS}(\textbf{d})|\mathcal{F}_{i,t-p-1}  \right ]=\mathbb{E}\left [ u_{i,t}^{RF}(\textbf{d})|\mathcal{F}_{i,t-p-1}  \right ]=0$ for all $\textbf{d}\in\left \{ 0,1 \right \}^{p+1}$. Furthermore, define $u_{i,t}^{RF}(\textbf{d},\tilde{\textbf{d}};p)=u_{i,t}^{RF}(\textbf{d})+u^{RF}_{i,t}(\tilde{\textbf{d}})$ and $u_{i,t}^{FS}(\textbf{d},\tilde{\textbf{d}};p)$ analogously. It follows that $\mathbb{E}\left [ u^{RF}_{i,t}(\textbf{d},\tilde{\textbf{d}};p)|\mathcal{F}_{i,t-p-1}  \right ]=\mathbb{E}\left [ u^{FS}_{i,t}(\textbf{d},\tilde{\textbf{d}};p)|\mathcal{F}_{i,t-p-1}  \right ]=0$.

Now let's consider the variance. I prove the result by induction for the reduced form and first stage of each potential outcome separately, and build on it to derive the properties of the estimator.

First, let's consider the case when $p=0$ and $D_{i,t}=1$ (the case for $D_{i,t}=0$ is analogous). For the first stage, it's been proved earlier that

\begin{align*}
    &\mathbb{V}\left [ u_{i,t}^{FS}(1)|\mathcal{F}_{i,t-p-1}  \right ]=\frac{D_{i,t}^{2}(1)}{\pi_{i,t}(1)}+\frac{D_{i,t}^{2}(0)}{\pi_{i,t}(0)}-(D_{i,t}(1)-D_{i,t}(0))^{2}\\&=\left ( \gamma_{i,t}^{FS}(1) \right )^{2}-(D_{i,t}(1)-D_{i,t}(0))^{2}=\left ( \sigma_{i,t}^{FS}(1) \right )^{2}
\end{align*}

Now let's consider the reduced form. Using a reasoning similar from the one in the lag-0 case, we obtain

\begin{align*}
    &\mathbb{V}\left [ u_{i,t}^{RF}(1)|\mathcal{F}_{i,t-p-1}  \right ]=\left ( Y_{i,t}(d^{obs}_{i,1:t-1},1)D_{i,t}(1) \right )^{2}\mathbb{V}\left [ W_{i,t}(1) \right ]+\left ( Y_{i,t}(d^{obs}_{i,1:t-1},1)D_{i,t}(0) \right )^{2}\mathbb{V}\left [ W_{i,t}(0) \right ]\\&-2Y_{i,t}(d^{obs}_{i,1:t-1},1)D_{i,t}(1)Y_{i,t}(d^{obs}_{i,1:t-1},1)D_{i,t}(0)\mathbb{C}ov\left [ W_{i,t}(1),W_{i,t}(0) \right ]\\&=\frac{\left ( Y_{i,t}(d^{obs}_{i,1:t-1},1)D_{i,t}(1) \right )^{2}}{\pi_{i,t}(1)}+\frac{\left ( Y_{i,t}(d^{obs}_{i,1:t-1},1)D_{i,t}(0) \right )^{2}}{\pi_{i,t}(0)}\\&-(Y_{i,t}(d^{obs}_{i,1:t-1},1)D_{i,t}(1)-Y_{i,t}(d^{obs}_{i,1:t-1},1)D_{i,t}(0))^{2}\\&=\left ( \gamma_{i,t}^{RF}(1) \right )^{2}-(Y_{i,t}(d^{obs}_{i,1:t-1},1)D_{i,t}(1)-Y_{i,t}(d^{obs}_{i,1:t-1},1)D_{i,t}(0))^{2}=\left ( \sigma_{i,t}^{RF}(1) \right )^{2}
\end{align*}

Now, I proceed by deriving the expression for a general $p+1$, assuming that the result holds for $p$. For the first stage, 

\begin{align*}
    &\mathbb{V}\left [ u_{i,t}^{FS}(\textbf{d})|\mathcal{F}_{i,t-p-1}  \right ]=\mathbb{V}\left [ \Delta^{p+2}\left (\mathbf{1}\left \{ D_{i,t-p-1:t}(\textbf{z})=\textbf{d} \right \}(W_{i,t-p-1}(\textbf{z})-1)  \right ) \right ]\\&=\mathbb{V} [ \Delta^{p+1}\left (\mathbf{1}\left \{ D_{i,t-p-1:t}(1,\textbf{z})=\textbf{d} \right \}(W_{i,t-p-1}(1,\textbf{z})-1)  \right ) \\&-\Delta^{p+1}\left (\mathbf{1}\left \{ D_{i,t-p-1:t}(0,\textbf{z})=\textbf{d} \right \}(W_{i,t-p-1}(0,\textbf{z})-1)  \right ) ]\\&=\mathbb{V}\left [ \Delta^{p+1}\left (\mathbf{1}\left \{ D_{i,t-p-1:t}(1,\textbf{z})=\textbf{d} \right \}(W_{i,t-p-1}(1,\textbf{z})-1)  \right )  \right ]\\&+\mathbb{V}\left [ \Delta^{p+1}\left (\mathbf{1}\left \{ D_{i,t-p-1:t}(0,\textbf{z})=\textbf{d} \right \}(W_{i,t-p-1}(0,\textbf{z})-1)  \right )  \right ]\\&-2 \Delta^{p+1}\left (\mathbf{1}\left \{ D_{i,t-p-1:t}(1,\textbf{z})=\textbf{d} \right \}\right)\Delta^{p+1}\left (\mathbf{1}\left \{ D_{i,t-p-1:t}(0,\textbf{z})=\textbf{d} \right \}\right)\mathbb{C}ov\left [ W_{i,t-p-1}(1,\textbf{z}),W_{i,t-p-1}(0,\textbf{z}) \right ]\\&=\sum_{\textbf{z}\in\left \{ 0,1 \right \}^{p+1}}\frac{\mathbf{1}\left \{ D_{i,t-p-1}(1,\textbf{z}) \right \}=\textbf{d}}{\pi_{i,t-p-1}(1,\textbf{z})}-\Delta^{p+1}\left (  D_{i,t-p-1}(1,\textbf{z})\right )^{2}\\&+\sum_{\textbf{z}\in\left \{ 0,1 \right \}^{p+1}}\frac{\mathbf{1}\left \{ D_{i,t-p-1}(0,\textbf{z}) \right \}=\textbf{d}}{\pi_{i,t-p-1}(0,\textbf{z})}-\Delta^{p+1}\left (  D_{i,t-p-1}(0,\textbf{z}_{-})\right )^{2}\\&+2\Delta^{p+1}\left (D_{i,t-p-1}(1,\textbf{z})  \right )\Delta^{p+1}\left ( D_{i,t-p-1}(0,\textbf{z}) \right )\\&=\sum_{\textbf{z}\in\left \{ 0,1 \right \}^{p+2}}\frac{\mathbf{1}\left \{ D_{i,t-p-1}(\textbf{z}) \right \}=\textbf{d}}{\pi_{i,t-p-1}(\textbf{z})}\\&-\left ( \Delta^{p+2}\left ( \mathbf{1}\left \{ D_{i,t-p-1}(1,\textbf{z}) \right \}=\textbf{d} \right )-\Delta^{p+2}\left ( \mathbf{1}\left \{ D_{i,t-p-1}(0,\textbf{z}) \right \}=\textbf{d} \right ) \right )^{2}\\&=\left ( \gamma_{i,t}^{FS}(\textbf{d}) \right )^{2}-\left ( \Delta^{p+2}\left ( \mathbf{1}\left \{ D_{i,t-p-1}(\textbf{z}) \right \}=\textbf{d} \right ) \right )^{2}\\&=\left ( \sigma_{i,t}^{FS}(\textbf{d}) \right )^{2}
\end{align*}

Analogously, for the reduced form it can be shown that

\begin{align*}
    &\mathbb{V}\left [ u_{i,t}^{RF}(\textbf{d})|\mathcal{F}_{i,t-p-1}  \right ]=\sum_{\textbf{z}\in\left \{ 0,1 \right \}^{p+2}}\frac{\left ( Y_{i,t}(d^{obs}_{i,1:t-p-2},\textbf{d})\mathbf{1}\left \{ D_{i,t-p-1:t}(\textbf{z}) =\textbf{d}\right \} \right )^{2}}{\pi_{i,t-p-1}(\textbf{z})}\\&-\left ( \Delta^{p+2}\left ( Y_{i,t}(d^{obs}_{i,1:t-p-2},\textbf{d})\mathbf{1}\left \{ D_{i,t-p-1:t}(\textbf{z}) =\textbf{d}\right \} \right ) \right )^{2}\\&=\left ( \gamma_{i,t}^{RF}(\textbf{d}) \right )^{2}-\left ( \Delta^{p+2}\left ( Y_{i,t}(d^{obs}_{i,1:t-p-2},\textbf{d})\mathbf{1}\left \{ D_{i,t-p-1:t}(\textbf{z}) =\textbf{d}\right \} \right ) \right )^{2}=\left ( \sigma_{i,t}^{RF}(\textbf{d}) \right )^{2}
\end{align*}

From the results above, we have 
\begin{align*}
    &\mathbb{V}\left [ u_{i,t}(\textbf{d})|\mathcal{F}_{i,t-p-1}  \right ]=g^{'}\begin{pmatrix}
\left ( \sigma_{i,t}^{RF}(\textbf{d}) \right )^{2}\\ 
\left ( \sigma_{i,t}^{FS}(\textbf{d}) \right )^{2}
\end{pmatrix}g=\left ( \sigma_{i,t}(\textbf{d}) \right )^{2}\\&\approx \frac{\left ( \sigma_{i,t}^{RF}(\textbf{d}) \right )^{2}}{\left ( m_{i,t}^{FS}(\textbf{d}) \right )^{2}}+\frac{\left ( m_{i,t}^{RF}(\textbf{d}) \right )^2\left ( \sigma_{i,t}^{FS}(\textbf{d}) \right )^{2}}{\left ( m_{i,t}^{FS}(\textbf{d}) \right )^{4}}-\frac{2m_{i,t}^{RF}(\textbf{d})\mathbb{C}ov\left ( \widehat{m}_{i,t}^{RF}(\textbf{d}),\widehat{m}_{i,t}^{FS}(\textbf{d}) \right )}{\left ( m_{i,t}^{FS}(\textbf{d}) \right )^{3}}
\end{align*}

Where $g$ is the gradient of $h(x,y)=x/y$ evaluated at $(m_{i,t}^{RF}(\textbf{d}),m_{i,t}^{FS}(\textbf{d}))$ 

The variance for $\widehat{\tau}_{i,t}(\textbf{d},\tilde{\textbf{d}};p)$ is simply $\left ( \sigma_{i,t}(\textbf{d}) \right )^{2}+\left ( \sigma_{i,t}(\tilde{\textbf{d}}) \right )^{2}$, which concludes the proof.

\section*{Appendix C - Causal Decomposition of the Period-Specific Wald Estimand}

I focus on the case of $t=2$, which is relevant for the results of the Monte Carlo simulation. Define $N_{2}^{(\textbf{z})}=\sum_{i=1}^{N}\mathbf{1}\left\{Z_{i,2}=\textbf{z}\right\}$ and $\omega_{i,2}=\mathbb{P}\left(Z_{i,2}=1\right)$. The period-specific Wald estimand for $t=2$ is

\begin{equation*}
    \beta_{2}^{2SLS}=\frac{\beta_{2}^{RF}}{\beta_{2}^{FS}}=\frac{\mathbb{E}\left [ \frac{1}{N_{2}^{(1)}}\sum_{i=1}^{N}Z_{i,2}Y_{i,2}-\frac{1}{N_{2}^{(0)}}\sum_{i=1}^{N}(1-Z_{i,2})Y_{i,2} \right ]}{\mathbb{E}\left [ \frac{1}{N_{2}^{(1)}}\sum_{i=1}^{N}Z_{i,2}D_{i,2}-\frac{1}{N_{2}^{(0)}}\sum_{i=1}^{N}(1-Z_{i,2})D_{i,2} \right ]}
\end{equation*}

I begin with the first stage. Under Assumptions 6-10,

\begin{align*}
    &\beta_{2}^{FS}=\mathbb{E}\left [ \frac{1}{N_{2}^{(1)}}\sum_{i=1}^{N}Z_{i,2}D_{i,2}-\frac{1}{N_{2}^{(0)}}\sum_{i=1}^{N}(1-Z_{i,2})D_{i,2} \right ]\\&= \frac{1}{N_{2}^{(1)}}\sum_{i=1}^{N}\omega_{i,2}D_{i,2}(1)-\frac{1}{N_{2}^{(0)}}\sum_{i=1}^{N}(1-\omega_{i,2})D_{i,2}(0)\\&= \frac{1}{N_{2}^{(1)}}\sum_{i\in C_{2}}\omega_{i,2}+\frac{1}{N_{2}^{(1)}}\sum_{i=1}^{N}\omega_{i,2}D_{i,2}(0)-\frac{1}{N_{2}^{(0)}}\sum_{i=1}^{N}(1-\omega_{i,2})D_{i,2}(0)\\&= \frac{1}{N_{2}^{(1)}}\sum_{i\in C_{2}}\omega_{i,2}+\frac{NN}{N_{2}^{(1)}N_{2}^{(0)}}\left ( \frac{1}{N}\sum_{i=1}^{N}\left ( \omega_{i,2}-\frac{N_{2}^{(1)}}{N}  \right )D_{i,2}(0) \right )\\&= \frac{1}{N_{2}^{(1)}}\sum_{i\in C_{2}}\omega_{i,2}+\frac{NN}{N_{2}^{(1)}N_{2}^{(0)}}\mathbb{C}ov\left ( \omega_{i,2},D_{i,2}(0) \right )\\&= \frac{1}{N_{2}^{(1)}}\sum_{i\in C_{2}}\omega_{i,2}
\end{align*}

For the reduced form, we have

{\small \begin{align*}
    &\mathbb{E}\left [ \frac{1}{N_{2}^{(1)}}\sum_{i=1}^{N}Z_{i,2}Y_{i,2}- \frac{1}{N_{2}^{(0)}}\sum_{i=1}^{N}(1-Z_{i,2})Y_{i,2} \right ]\\&=\mathbb{E}\left [ \frac{1}{N_{2}^{(1)}}\sum_{i=1}^{N}Z_{i,2}Y_{i,2}(D_{i,1}(Z_{i,1}),D_{i,2}(1))- \frac{1}{N_{2}^{(0)}}\sum_{i=1}^{N}(1-Z_{i,2})Y_{i,2}(D_{i,1}(Z_{i,1}),D_{i,2}(0)) \right ]\\&=\mathbb{E}\left [ \frac{1}{N_{2}^{(1)}}\sum_{i=1}^{N}Z_{i,1}Z_{i,2}Y_{i,2}(D_{i,1}(1),D_{i,2}(1))+(1-Z_{i,1})Z_{i,2}Y_{i,2}(D_{i,1}(0),D_{i,2}(1)) \right ]\\&-\mathbb{E}\left [ \frac{1}{N_{2}^{(0)}}\sum_{i=1}^{N}Z_{i,1}(1-Z_{i,2})Y_{i,2}(D_{i,1}(1),D_{i,2}(0))+(1-Z_{i,1})(1-Z_{i,2})Y_{i,2}(D_{i,1}(0),D_{i,2}(0)) \right ]\\&=\frac{1}{N_{2}^{(1)}}\sum_{i=1}^{N}\pi_{i,1:2}(1,1)Y_{i,2}(D_{i,1}(1),D_{i,2}(1))+\sum_{i=1}^{N}\pi_{i,1:2}(0,1)Y_{i,2}(D_{i,1}(0),D_{i,2}(1))\\&-\left(\frac{1}{N_{2}^{(0)}}\sum_{i=1}^{N}\pi_{i,1:2}(1,0)Y_{i,2}(D_{i,1}(1),D_{i,2}(0))+\sum_{i=1}^{N}\pi_{i,1:2}(0,0)Y_{i,2}(D_{i,1}(0),D_{i,2}(0))\right)\\&=\frac{1}{N_{2}^{(1)}}\sum_{i=1}^{N}\pi_{i,1:2}(1,1)\left\{Y_{i,2}(D_{i,1}(1),D_{i,2}(1))-Y_{i,2}(D_{i,1}(1),D_{i,2}(0))+Y_{i,2}(D_{i,1}(1),D_{i,2}(0)) \right\}\\&+\frac{1}{N_{2}^{(1)}}\sum_{i=1}^{N}\pi_{i,1:2}(0,1)\left\{Y_{i,2}(D_{i,1}(0),D_{i,2}(1))-Y_{i,2}(D_{i,1}(0),D_{i,2}(0))+Y_{i,2}(D_{i,1}(0),D_{i,2}(0)) \right\}\\&-\frac{1}{N_{2}^{(0)}}\sum_{i=1}^{N}\pi_{i,1:2}(1,0)\left\{Y_{i,2}(D_{i,1}(1),D_{i,2}(0))-Y_{i,2}(D_{i,1}(0),D_{i,2}(0))+Y_{i,2}(D_{i,1}(0),D_{i,2}(0)) \right\}\\&-\frac{1}{N_{2}^{(0)}}\sum_{i=1}^{N}\pi_{i,1:2}(0,0)Y_{i,2}(D_{i,1}(0),D_{i,2}(0))\\&=\frac{1}{N_{2}^{(1)}}\sum_{i=1}^{N}\pi_{i,1:2}(1,1)\left\{ Y_{i,2}(D_{i,1}(1),D_{i,2}(1))-Y_{i,2}(D_{i,1}(1),D_{i,2}(0)) + Y_{i,2}(D_{i,1}(1),D_{i,2}(0))-Y_{i,2}(D_{i,1}(0),D_{i,2}(0))\right\}\\&+\frac{1}{N_{2}^{(1)}}\sum_{i=1}^{N}\pi_{i,1:2}(0,1)\left\{ Y_{i,2}(D_{i,1}(0),D_{i,2}(1))-Y_{i,2}(D_{i,1}(0),D_{i,2}(0)) \right\}+\frac{1}{N_{2}^{(1)}}\sum_{i=1}^{N}\omega_{i,2}Y_{i,2}(D_{i,1}(0),D_{i,2}(0))\\&-\frac{1}{N_{2}^{(0)}}\sum_{i=1}^{N}\pi_{i,1:2}(1,0)\left\{ Y_{i,2}(D_{i,1}(1),D_{i,2}(0))-Y_{i,2}(D_{i,1}(0),D_{i,2}(0)) \right\}+\frac{1}{N_{2}^{(0)}}\sum_{i=1}^{N}(1-\omega_{i,2})Y_{i,2}(D_{i,1}(0),D_{i,2}(0))\\&=\frac{1}{N_{2}^{(1)}}\sum_{i\in C_{2}}\omega_{i,2}\tau_{i,2}(1,0;0)+\frac{1}{N_{2}^{(1)}}\sum_{i\in C_{1}}\pi_{i,1:2}(1,1)\tau_{i,2}(\left\{ 1,D_{i,2}(0)\right\},\left\{ 0,D_{i,2}(0)\right\};1)\\&-\frac{1}{N_{2}^{(0)}}\sum_{i\in C_{1}}\pi_{i,1:2}(1,0)\tau_{i,2}(\left\{ 1,D_{i,2}(0)\right\},\left\{ 0,D_{i,2}(0)\right\};1)+\frac{NN}{N_{2}^{(1)}N_{2}^{(0)}}\mathbb{C}ov\left ( \omega_{i,2},Y_{i,2}(D_{i,1}(0),D_{i,2}(0)) \right )
\end{align*}}

and thus, we conclude that

\begin{align*}
    &\beta_{2}^{RF}=\frac{1}{N_{2}^{(1)}}\sum_{i\in C_{2}}\omega_{i,2}\tau_{i,2}(1,0;0)+\frac{1}{N_{2}^{(1)}}\sum_{i\in C_{1}}\pi_{i,1:2}(1,1)\tau_{i,2}(\left\{ 1,D_{i,2}(0)\right\},\left\{ 0,D_{i,2}(0)\right\};1)\\&-\frac{1}{N_{2}^{(0)}}\sum_{i\in C_{1}}\pi_{i,1:2}(1,0)\tau_{i,2}(\left\{ 1,D_{i,2}(0)\right\},\left\{ 0,D_{i,2}(0)\right\};1),\\&\beta_{2}^{FS}=\frac{1}{N_{2}^{(1)}}\sum_{i\in C_{2}}\omega_{i,2}
\end{align*}

\end{document}